\documentclass{article}

\usepackage{arxiv}
\usepackage{xcolor}
\usepackage[utf8]{inputenc} 
\usepackage[T1]{fontenc}    
\usepackage{hyperref}       
\usepackage{url}            
\usepackage{booktabs}       
\usepackage{amsfonts}       
\usepackage{nicefrac}       
\usepackage{microtype}      
\usepackage{lipsum}		
\usepackage{graphicx}
\usepackage[numbers]{natbib}
\usepackage{doi}
\usepackage{amsmath}
\usepackage{subcaption}
\usepackage{float}

\newcommand{\Sup}[1]{S^{\uparrow}\!\left[#1\right]}
\newcommand{\Sdown}[1]{S^{\downarrow}\!\left[#1\right]}
\newcommand{\SItoE}{\Sup{W_{I\to E}}}
\newcommand{\SrI}{\Sup{r_{I}}}

\newcommand{\SEE}{\Sdown{W_{E\to E}}}
\newcommand{\SrE}{\Sdown{r_{E}}}
\newcommand{\SEI}{\Sup{W_{I\to E}}}

\newcommand{\SWup}{\Sup{W}}
\newcommand{\SWdown}{\Sdown{W}}

\newcommand{\SrUp}{\Sup{r}}
\newcommand{\SrDown}{\Sdown{r}}

\newcommand{\remove}[1]{}

\title{Modelling chronic stress as excitation–inhibition perturbation in recurrent working-memory networks}

\author{ \href{https://orcid.org/0009-0006-6640-9273}{\includegraphics[scale=0.06]{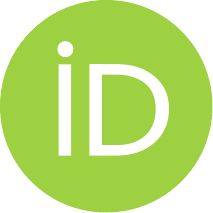}\hspace{1mm}Mauricio A. Diaz}\\
	Leibniz Institute for Resilience Research (LIR)\\
	Institute for Quantitative and Computational Biosciences (IQCB), Johannes Gutenberg University\\
	 Mainz, Germany \\
	\texttt{mauricio.a-diaz@lir-mainz.de} \\
	\And
	\href{https://orcid.org/0009-0004-8166-1675}{\includegraphics[scale=0.06]{orcid.pdf}\hspace{1mm}Manuela A.~Beyer} \\
	Leibniz Institute for Resilience Research (LIR)\\
	Institute for Quantitative and Computational Biosciences (IQCB), Johannes Gutenberg University\\
	 Mainz, Germany \\
	\texttt{manuela.beyer@lir-mainz.de} \\
	\And
	\href{https://orcid.org/0000-0003-0283-3471}{\includegraphics[scale=0.06]{orcid.pdf}\hspace{1mm}Janina~Hesse} \\
	Leibniz Institute for Resilience Research (LIR)\\
	University Medical Center of the Johannes Gutenberg University\\
	Institute for Quantitative and Computational Biosciences (IQCB), Johannes Gutenberg University\\
	 Mainz, Germany \\
	\texttt{janina.hesse@lir-mainz.de} \\
}

\renewcommand{\shorttitle}{Chronic stress as E/I perturbation in RNNs}

\hypersetup{
pdftitle={Modelling chronic stress as excitation–inhibition perturbation in recurrent working-memory networks},
pdfsubject={q-bio.NC, q-bio.QM},
pdfauthor={Mauricio A. Diaz, Manuela A. Beyer, Janina Hesse},
pdfkeywords={Chronic stress, Excitatory-inhibitory balance, Recurrent neural networks, Working memory, Prefrontal cortex, Resilience, Network simulation},
}

\begin{document}
\maketitle

\begin{abstract}
Chronic stress can turn an organism's stress response maladaptive, with consequences including structural brain changes, behavioral inflexibility and impaired working memory. At the cellular level, chronic stress shifts the excitatory-inhibitory (E/I) balance of prefrontal pyramidal neurons toward inhibitory dominance, yet the mechanisms underlying these alterations, as well as their effect on network performance, are still unknown. 
Here, we use recurrent excitatory–inhibitory networks as minimal circuit-level models to isolate a chronic-stress-associated increase in inhibition and examine its consequences for inhibitory dominance, excitatory hypofunction, and computation. We then ask how stress adaptation preserves function and what computational costs such adaptation entails. 
We model the circuit-level effects of stress as selective stochastic strengthening of inhibitory synapses onto excitatory neurons. Applying this stress perturbation to neural networks trained on a working-memory task, task performance drops, as would be expected under chronic stress, and both excitatory and inhibitory drive collapse. In contrast, networks trained under strengthened inhibition preserve task performance and develop inhibitory dominance visible as reduced excitatory drive alongside maintained or increased inhibition. These resilient networks constrain their recurrent dynamics under stress close to their unstressed trajectories, stabilises an energetic proxy, and produces a sparser, less reciprocally connected circuit. These stress adaptations, however, decrease performance of resilient networks compared to networks trained without stress for tasks requiring longer working-memory. This resilience-generalisation trade-off persists across network size and stress magnitude used during training. A comparison of eight candidate perturbations on synaptic strength or neuronal excitability confirms that strengthened inhibition best reproduces the joint circuit-level signatures of stress among those tested. Our results suggest that adaptation on the network level can preserve familiar computations under stress, yet stress adaptation specialises circuits such that flexibility outside the trained regime is limited.
\end{abstract}




\section{Introduction}

Chronic stress impairs prefrontal-dependent cognitive processes in both humans and rodents.
Across experimental and clinical studies, chronic stress reduces working memory and cognitive flexibility, while promoting rigid,
habit-driven strategies~\cite{Arnsten2009-pn,Kim2023-yu,Dias-Ferreira2009-nh,Girotti2024-vt}.
These functional changes are accompanied by structural remodelling of
prefrontal pyramidal neurons~\cite{McEwen2016-wi}. A central candidate
mechanism is disruption of the excitatory/inhibitory (E/I) balance~\cite{Page2019-xg}. E/I balance is thought to support
stable yet flexible information processing in the prefrontal cortex (PFC), and
its disruption is implicated in depression and anxiety disorders
~\cite{Sears2021-lz,Kirischuk2022-ml}.
Converging evidence indicates that prolonged stress shifts medial PFC (mPFC)
circuitry toward inhibitory dominance, with most studies pointing towards increased inhibitory
drive onto pyramidal neurons: in mice, chronic stress shifts the synaptic E/I balance of layer 5/6 pyramidal neurons toward inhibition~\cite{Rodrigues2024-ry}, and in rats, chronic stress strengthens inhibitory synapses, both by enhanced presynaptic
GABA release and increased GABAergic contacts onto infralimbic pyramidal cells. Rats furthermore show a selective loss of glucocorticoid receptors in parvalbumin (PV)
interneurons~\cite{McKlveen2016-px}, while in mice synaptic drive onto PV interneurons is unaffected~\cite{Rodrigues2024-ry}. 

Spiking circuit models, mean-field analyses, and recurrent networks have shown
that E/I balance shapes evidence accumulation, attractor stability, and
competition between representations, with inhibition contributing actively to
persistent activity rather than acting purely as a suppressive force
~\cite{Lam2022-ly,Roach2023-ti}. Task-trained recurrent networks are by now a
standard framework for studying working-memory computation under biologically
grounded constraints such as Dale's principle
~\cite{Song2016-qj,Ingrosso2019-mr,Masse2019-xj,Cueva2021-hs,Piwek2023-sg}, and
recent work has begun connecting stress and neuromodulation to recurrent
computation~\cite{Du2025-ri,Tsuda2026-tl}, alongside studies of noise injection
and weight perturbation as routes to regularisation and robustness
~\cite{Lim2021-yy,Wu2020-mt,Rungratsameetaweemana2025-cc}. What has not been
tested is whether a E/I perturbation biologically motivated by chronic stress reproduces the
experimental signatures of stress-induced prefrontal dysfunction, and which
cost adapting to such a perturbation implies.

\begin{figure}[ht!]
    \centering
    \includegraphics[width=0.98\linewidth]{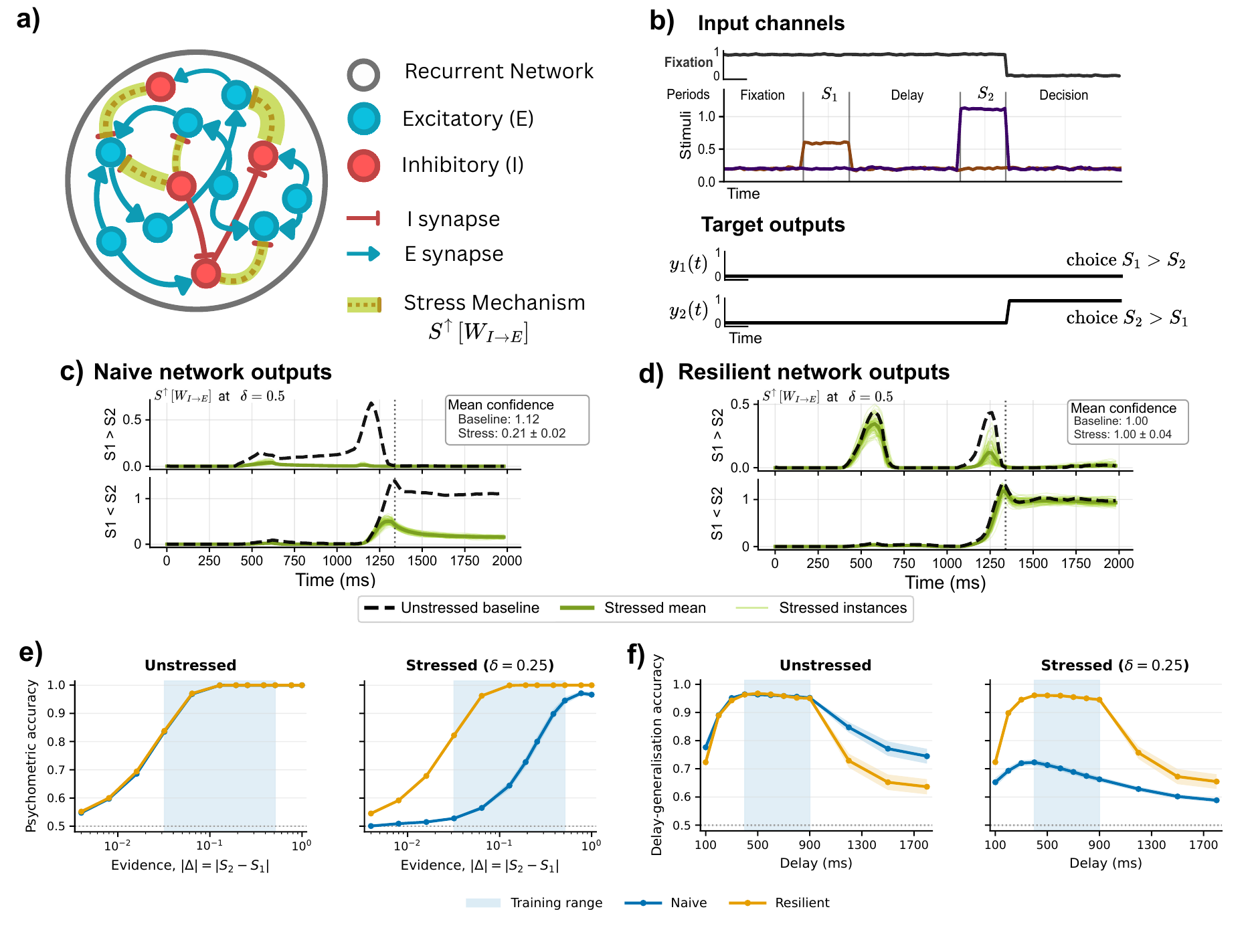}
    \caption{%
    \textbf{Resilience training preserves comparison performance under
    inhibitory-to-excitatory stress but reduces stimulus delay generalisation.}
    \textbf{(a)} Dale-constrained recurrent network with excitatory (blue, $E$) and
    inhibitory (red, $I$) units. The canonical stress operator $\SEI$
    increases $I\to E$ synapses in the mean by 50\% (stress magnitude $\delta=0.5$, green).
    \textbf{(b)} Delayed parametric-comparison task. The network receives distinct inputs $S_1$
    and $S_2$, separated by a delay period, and reports the
    larger stimulus after the fixation signal (Fixation) is released.
    \textbf{(c,d)} Output trajectories from representative naive (c) and resilient
    networks (d) for one $S_1<S_2$ trial at $\delta=0.5$. Dashed black
    traces show baseline, light green traces show individual stress instances,
    and thick green traces show their mean. Decision epoch starts at the dotted vertical lines.
    \textbf{(e,f)} Psychometric accuracy without stress and under test-time stress with
    $\delta=0.25$ as a function of evidence
    $|\Delta|=|S_2-S_1|$ (e), and of the delay between $S_1$ and $S_2$ (f). Blue shading marks the training range. Curves show
    means over 50 independently trained networks per class; bands show the standard error of the mean (SEM).}
    \label{fig:figure1}
\end{figure}

Here, we study chronic stress \emph{in silico} using recurrent neural networks
(RNNs) constrained by Dale's principle, which enforces distinct excitatory and
inhibitory populations~\cite{Jarne2024-mm,Barranca2022-aj}. The neurons are modelled as rate units, such that changes in synaptic weights represent both presynaptic and postsynaptic
contributions, including an increased number of
release sites, a higher release probability, and a stronger postsynaptic response. 
We model chronic
stress as a stochastic multiplicative operator $S$ acting on the recurrent
weights from inhibitory neurons to excitatory neurons, $|W_{I\to E}|$, with a mean magnitude $\delta$ setting the directional shift and a
relative dispersion $\sigma_{\mathrm{stress}}$ setting the spread across
synapses.  This stress operator 
$\SItoE$ induces a directional and heterogeneous increase in inhibition, while remaining agnostic about the molecular
process, thereby providing a natural minimal abstraction of the experimental results. 
We compare $\SItoE$ against alternative operators targeting either recurrent
excitation, all recurrent synapses, or population gain, and find that only
$\SItoE$ reproduces the inhibitory dominance observed experimentally. Because E/I balance has been shown to modulate working memory~\cite{Wang1999-kl,Compte2000-gw}, we evaluate cognitive performance on a working
memory task. For the task, the network reports the higher of two stimuli $S_1$ and
$S_2$, presented with a variable delay
~\cite{Song2016-qj,Yang2019-sq,Yang2020-hj}, see
Fig.~\ref{fig:figure1}. The task isolates three core functions:
encoding sensory information, maintaining it across the delay between $S_1$ and
$S_2$, and making a decision from the integrated evidence. 

Networks trained solely on the working memory task, called \emph{naive networks}
in the following, are highly sensitive to the stress operator. To model
adaptation to chronic stress, we introduced a \emph{resilience training} motivated by the
observation that animals are most resilient against stress when subjected to a
moderate amount of stress during development (stress inoculation,
~\cite{kalisch_neurobiology_2024}): we trained networks on the memory task while
simultaneously employing the stress operator. These networks maintained task
performance under stress, and we call them, following the stress-resilience
literature~\cite{Nestler2024-jj}, \emph{resilient networks}. Resilience in this
sense denotes maintenance of function under sustained adversity; it is defined
relative to a specified outcome and does not imply preserved performance on other
measures. Note that, from a network-theoretical point of view, our resilient
networks might be seen, due to their lack of plasticity after training, rather as
robust than resilient to the stress operator.

Our contributions are threefold. First, we show that resilience training
preserves task performance under sustained inhibitory perturbation but produces
a loss of delay generalisation that is not encoded anywhere in the training
objective, an emergent behavioural signature aligned with stress-induced
inflexibility; this resilience--generalisation trade-off persists across network size and the
magnitude of training stress, indicating that it is not
simply resolved by additional capacity. Second, we characterise how resilience is
implemented dynamically, energetically, and structurally: resilient networks
constrain their activity trajectories 
to remain closer to the unstressed
baseline geometry, operate from a lower and more stable metabolic-energy
regime, and settle on a sparser, less reciprocally connected recurrent
topology. Third, we confirm that this account does not rest on our chosen stress-mimicking perturbation: systematically comparing our biologically motivated,
Dale-preserving stochastic stress operator against a family of alternative
perturbations, only inhibitory upscaling onto excitatory cells reproduces the
inhibitory dominance reported experimentally, supporting it as a promising
minimal stress operator.
Together, these findings suggest that chronic stress-induced inflexibility can
be the by-product of a circuit specialising to keep functioning under sustained
stress, rather than an independently emerging deficit.

\section{Preliminaries}
\paragraph{Excitatory-inhibitory recurrent network model and working memory task.}
Following the work by Song \textit{et al.}~\cite{Song2016-qj} and others~\cite{Rajakumar2021-uh, Ingrosso2019-mr, Yang2019-sq, Yang2020-hj}, our networks consist of 200 rate-based neurons, $80\%$ excitatory and $20\%$ inhibitory, following Dale's principle~\cite{Dale1935-yk}. Full details are
given in the Methods~\ref{sec:network_model} and a
representative network schematic is shown in Fig.~\ref{fig:figure1}(a).
Neurons are all-to-all connected according to the connectivity matrix $W_{\mathrm{rec}}$, consisting of submatrices $W_{E \to E}$, $W_{E \to I}$, $W_{I\to E}$, $W_{I \to I}$.
The activity of the neurons is denoted by $r_E$ for excitatory and $r_I$ for inhibitory populations. To evaluate the excitatory-inhibitory ratio, we introduce the excitatory drive $h_{E\to E}(t)$ as the sum of the recurrent input from excitatory to excitatory neurons, and the inhibitory drive $h_{I\to E}(t)$ accordingly, both defined in Eq.~\ref{eq:main_drives_E} in the Methods \ref{sec:network_model}.

The output of the network is decoded by the excitatory population,  $y(t) = W_{\mathrm{out}} r_E$, motivated by the predominance of long-range excitatory connections to  downstream areas~\cite{Harris2015-gh}.

We trained the networks on a delayed parametric-comparison task in which the
network reports which of two presented stimuli is larger. The network receives
three independent scalar input channels: two encode the stimuli $S_1$ and
$S_2$, separated by a variable delay $\tau$, and the third is a fixation input
that indicates the onset of the decision epoch. Fig.~\ref{fig:figure1}(b)
illustrates a representative trial. Before the decision epoch, both output
targets are zero. Immediately after $S_2$, the fixation input drops and the
two-dimensional target becomes $[1,0]$ for $S_1>S_2$ and $[0,1]$ for
$S_2>S_1$. The continuous-valued network output $\mathbf{y}(t)$ is trained
against these targets using a temporally weighted mean-squared error.

Trial difficulty is controlled by the stimulus difference
$\Delta = S_2 - S_1$, 
with small $|\Delta|$ producing ambiguous trials. Implementation details can be found in the Methods~\ref{sec:task_details}.
By sampling task difficulty $\Delta$ across a range of values we can measure how performance degrades as the
comparison becomes more ambiguous, yielding a psychometric curve, and how the internal
dynamics adapt to preserve decision accuracy across difficulty. The delay period lets us
probe parametric working memory directly, by examining how the network holds the first
stimulus $S_1$ in its recurrent activity and how well it generalises to out-of-distribution
delays.

\paragraph{Stress operators and target signatures.} \label{sec:targets}
Experimental studies of chronically stressed prefrontal cortex converge on two
circuit-level observations at pyramidal neurons: stronger inhibitory influence
and weaker excitatory output~\cite{McKlveen2016-px,Rodrigues2024-ry,Page2019-xg}.
We therefore use $\SEI$ as the primary stress operator. It selectively strengthens
inhibitory synapses onto excitatory units, $W_{I\to E}$, directly implementing the
first observation while allowing excitatory hypofunction to result from the network adaptation to stress. 
Stress magnitude $\delta$ sets the mean proportional scaling of
$W_{I\to E}$, drawn independently per synapse from a lognormal
distribution, and $\sigma_{\mathrm{stress}}$ sets its heterogeneity across
synapses (Methods, Section~\ref{sec:stress_mecs}).

We assessed $\SEI$ at circuit, population, and behavioural levels, and compared to experimental observations 
(Table~\ref{tab:targets}). Inhibitory dominance was quantified by a lower
$h_{E\to E}/|h_{I\to E}|$ drive ratio, and excitatory hypofunction by lower
$\bar r_E$ and $h_{E\to E}$. For the behavioural mapping, the broad phenotype of
stress-related cognitive impairment does not correspond a priori to a unique RNN
variable. Accuracy and MSE quantify task performance, whereas the
present results identify reduced accuracy at untrained, long delays between stimuli as a
model-level correlate of impaired cognitive flexibility, the third signature in
Table~\ref{tab:targets}. Three alternative stress operators consisting of population
targeted perturbations and four population-agnostic control operators were investigated
as a secondary specificity analysis (Fig.~\ref{fig:stress_operator_comparison});
their definitions and evaluation details are provided in the Supporting Information, S2 Text.
\begin{table}[ht]
\centering
\small
\renewcommand{\arraystretch}{1.3}
\begin{tabular}{|p{3.4cm}|p{5.4cm}|p{6.0cm}|}
\hline
\textbf{Signature} & \textbf{Experimental evidence} & \textbf{RNN operationalisation} \\
\hline
Inhibitory dominance at excitatory neurons
& Decreased sEPSC/sIPSC ratio in pyramidal neurons~\cite{Rodrigues2024-ry}; 
increased GABAergic contacts and release onto mPFC pyramidal neurons~\cite{McKlveen2016-px}
& Decreased E/I ratio of recurrent drives onto excitatory units:
$h_{E\to E}(t)\,/\,|h_{I\to E}(t)|$ \\
\hline
Excitatory hypofunction
& Reduced sEPSC frequency and amplitude~\cite{Rodrigues2024-ry}; diminished prefrontal influence on downstream targets~\cite{McKlveen2016-px}
& Decreased mean excitatory population activity $\bar r_E(t)$ and recurrent
excitatory drive $h_{E\to E}(t)$ \\
\hline
Reduced cognitive flexibility
& Chronic stress weakens working memory and behavioural flexibility and promotes
rigid, habit-like strategies~\cite{Arnsten2009-pn,Dias-Ferreira2009-nh,Girotti2024-vt}
& Lower task accuracy at out-of-distribution long delays ($>900~\mathrm{ms}$,
beyond the trained range). \\
\hline
\end{tabular}
\caption{\textbf{Biological signatures and their operationalisation in the RNN.}
The circuit-level signatures motivated $\SEI$ as the primary perturbation. The
behavioural signature is interpreted through the delay-generalisation result of
the present study; acute task accuracy and MSE are reported separately as direct
performance measures.}
\label{tab:targets}
\end{table}

\paragraph{Resilience training.}
Throughout this work we contrast networks trained on the task alone
(\emph{naive}) with networks trained while instances of $\SEI$ up to magnitude
$\delta_T=0.5$ were applied during training (\emph{resilient}; procedure in
Section~\ref{sec:stress_training}). Resilience here means maintenance of performance for the
trained task under the perturbation and does not imply generalisation to other
operators or task regimes.

\section{Results}

\subsection{Resilience training preserves task performance under inhibitory-to-excitatory stress}
\label{sec:results_performance}

We examined here the effects of the canonical stress operator $\SEI$, which increases inhibition of excitatory neurons, on network performance. Fig.~\ref{fig:figure1}a 
shows a representative trial in which application of the stress operator strongly attenuated the naive network's
correct output: mean decision confidence fell from $1.12$ at unstressed baseline to
$0.209\pm0.019$ across stress instances (Fig.~\ref{fig:figure1}c). In contrast, the resilient
network followed its baseline output, with confidence changing only
from $0.998$ to $0.958\pm0.044$ (Fig.~\ref{fig:figure1}d). This distinction
generalised across 50 independently trained networks per class
(mean$\pm$SEM throughout). Without stress, the psychometric curves were nearly
superimposed, with mean accuracies of $0.886\pm0.002$ and $0.888\pm0.002$
across evidence magnitudes for naive and resilient networks, respectively.
Under $\delta=0.25$ stress, resilient networks retained a similar psychometric
profile ($0.883\pm0.001$), whereas naive accuracy reduced to $0.714\pm0.008$.
This separation was evident across evidence strengths
(Fig.~\ref{fig:figure1}e): at $|\Delta|=0.032$,
accuracy under stress was $0.528\pm0.004$ in naive networks and
$0.822\pm0.006$ in resilient networks; at $|\Delta|=0.128$, it was
$0.644\pm0.013$ and $0.999\pm0.0002$, respectively. Thus, at stress magnitudes used during resilience
training, resilient networks largely preserved their evidence-dependent
responses, whereas naive networks did not.

Independent two-sided Welch $t$-tests showed no detectable class difference in
evidence-averaged accuracy between naive and resilient networks without stress (Holm-adjusted $p=0.553$), but a large difference under applied stress (naive-minus-resilient $\Delta=-0.169$, 95\% CI
$[-0.185,-0.153]$, Holm-adjusted $p<10^{-26}$). 
For each network, the stress
effect was calculated as stressed minus unstressed accuracy; this change due to stress also
differed between classes ($\Delta_{\mathrm{change}}=-0.167$, 95\% CI $[-0.184,-0.151]$,
Holm-adjusted $p<10^{-25}$). 
Comparisons of evidence-averaged accuracy, 
together with selected evidence-specific comparisons, are reported in 
Supplementary Table~\hyperref[tab:psychometric_class_stats]{S1}. Overall, resilience training strongly improves task performance under applied stress.

Better task performance in resilient networks came with a cost: resilience training introduced a
delay-generalisation trade-off (Fig.~\ref{fig:figure1}f). Averaged over the
trained delay range ($400$--$900\,\mathrm{ms}$), the two classes performed
similarly without stress (naive, $0.959\pm0.003$; resilient,
$0.959\pm0.003$). Under $\delta=0.25$ stress, as expected, accuracy fell to
$0.694\pm0.010$ in naive networks, but remained $0.955\pm0.002$ in resilient
networks (Fig.~\ref{fig:figure1}f, $400$--$900\,\mathrm{ms}$ delay). We observe a trade-off between resilience under stress and task generalisation without stress for long out-of-distribution delays: with applied stress, resilient networks outperformed
naive networks as expected ($0.606\pm0.008$ versus $0.695\pm0.024$), whereas without stress, the naive network gained considerable accuracy and outperformed resilient networks that remained at a similar performance level ($0.788\pm0.023$ versus $0.673\pm0.026$), see Fig.~\ref{fig:figure1}f, $1200$--$1800\,\mathrm{ms}$ delay.
Independent two-sided Welch $t$-tests found no detectable class difference
within the trained delay range without stress ($p=0.992$), whereas naive
networks were less accurate under stress ($p<10^{-30}$). At long delays, naive
networks were more accurate without stress ($p=0.00122$) and less accurate
under stress ($p=0.00104$). Effect sizes, confidence intervals, and test statistics are reported in Supplementary Table~\hyperref[tab:figure1_delay_stats]{S2}.

\begin{figure}[!ht]
    \centering
    \includegraphics[width=0.98\linewidth]{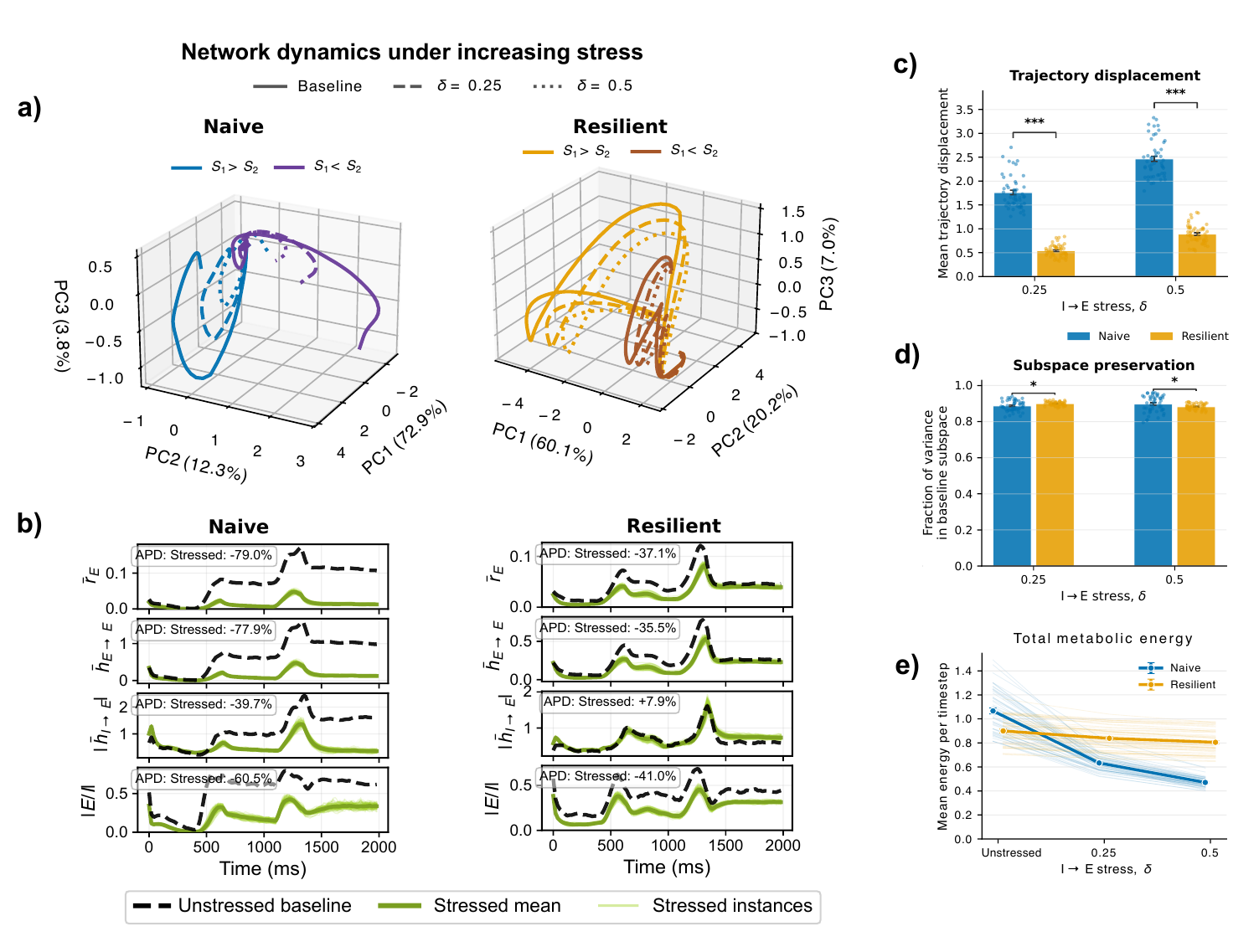}
    \caption{%
    \textbf{Resilience training constrains the dynamical and energetic response
    to $\mathbf{\SEI}$ stress.}
    \textbf{(a)} Recurrent trajectories from representative naive and resilient
    networks projected into a baseline PCA subspace for the two choices with
    $\Delta=\pm0.512$. Solid, dashed, and dotted lines denote baseline with $\delta=0$,
    $\delta=0.25$, and $\delta=0.5$, respectively.
    \textbf{(b)} Example-trial traces of mean excitatory population activity
    $\bar r_E$, recurrent excitatory drive $\bar h_{E\to E}$, absolute
    inhibitory drive onto excitatory units $|\bar h_{I\to E}|$, and the
    excitatory-to-inhibitory drive ratio $E/|I|$ at $\delta=0.5$. Dashed black
    traces show baseline, light green traces show individual stress instances,
    and thick green traces show their mean; insets report the signed average percentage difference (APD) to baseline.
    \textbf{(c)} Mean paired trajectory displacement in the fixed baseline PCA
    subspace, two-sided Mann–Whitney \textit{U}
tests naive versus
resilient, Holm-adjusted $p<10^{-16}$ at both stress levels.
    \textbf{(d)} Fraction of stressed-activity variance captured by same
    baseline PCA subspace (Holm-adjusted $p=0.045$ at both stress levels). Full test statistics are reported in Supplementary Table~\hyperref[tab:geometry_stats]{S3}.
    \textbf{(e)} Mean total metabolic energy per timestep at baseline (BL) with $\delta=0$ and under
    stress ($\delta\in\{0.25,0.5\}$).
    In \textbf{(c--e)}, points or light lines show individual networks
    and bars or dark lines show group means$\pm$SEM ($n=50$ per class).
    }
    \label{fig:figure2}
\end{figure}
\subsection{Resilience training stabilises recurrent dynamics}
\label{sec:recurrent_dynamics_geometry}

We next asked how the internal recurrent dynamics of naive and resilient networks gave 
rise to their differ behavioural responses under stress. For each of the 50 independently 
trained networks per class, principle component analysis (PCA) was fitted to its unstressed baseline recurrent 
activity, and both unstressed and stressed recurrent activity trajectories where projected into this joint space. Specifically, the smallest subspace explaining at least $90\%$ of the baseline 
variance was used for comparisons across stress levels. As a representative 
example from each class, we selected the network whose displacement and
subspace-preservation statistics were closest
to the center of the group's distribution, and projected its trajectories
for two trial choices onto the first three baseline principle components (PCs, Fig.~\ref{fig:figure2}a). 
Under stress, the naive network's trajectories collapsed, whereas the resilient 
network's displacement grew more gradually, preserving more of the baseline trajectory 
geometry as stress increased.

This pattern held across the full population of networks, resilience training maintained trajectories close to baseline even under stress; it significantly
limited trajectory displacement away from baseline (Fig.~\ref{fig:figure2}c). 
Even resilient networks at double stress magnitude ($\delta=0.5$) were less
displaced than naive networks at $\delta=0.25$ (two-sided Mann–Whitney test $U=7$, $p<10^{-16}$).

The fraction of stressed variance captured by the baseline subspace remained high 
in both groups 
(Fig.~\ref{fig:figure2}d). 
The significant but small decrease in subspace preservation in resilient networks suggests that resilience training does
not act by confining stressed activity to a different subspace, but by limiting
how far that activity moves within a shared baseline geometry, consistent with
the similarity of trajectories for resilient networks reported above.

Fig.~\ref{fig:figure2}b shows the population rates and drives during an
example trial for a representative network of each class, allowing us to
visualise how these cellular dynamics are affected by stress at test time. The naive network showed a broad collapse: $\bar r_E$ and
$\bar h_{E\to E}$ decreased by more than $75\%$, and
$|\bar h_{I\to E}|$ also fell by $40\%$ despite the upscaled inhibitory
weights. For the resilient networks the reductions in
$\bar r_E$ and $\bar h_{E\to E}$ were considerably smaller, less than 40\%, while
$|\bar h_{I\to E}|$ increased by $8\%$. Consequently, the E/I ratio $E/|I|$ decreased in
both networks (naive $61\%$ and resilient $41\%$),
but for different reasons, naive networks overall loose drive while resilient networks reduce excitatory drive but maintain inhibitory drive. This shift is the result of the resilience training's
adaptation to $\SEI$ stress: In naive networks, we observe a loss of recurrent drive from excitatory and inhibitory populations under stress, stress-increased inhibition reduces excitatory activity so strongly that also the inhibitory population cannot maintain baseline activity. In contrast, resilient networks actively preserve inhibitory drive onto
excitatory units, see $|\bar h_{I\to E}|$ in Fig.~\ref{fig:figure2}b, and this results in a reduction of $\bar r_E$ and $\bar h_{E\to E}$ that is sufficient to maintain inhibitory population activity (Fig.~\ref{fig:figure2}b), overall reflecting an adaptive strategy against stress.

The above reported distinction resulting from resilience training was consistent across the 50 independently trained networks
per class. For each network, stressed activity and drive were expressed as a
signed percentage change from its paired unstressed baseline. At
$\delta=0.5$, the median changes in naive networks were $-77.19\%$ for
$\bar r_E$, $-60.18\%$ for $\bar r_I$, $-76.48\%$ for
$\bar h_{E\to E}$, and $-39.27\%$ for $|\bar h_{I\to E}|$; all differed from
zero after Benjamini--Hochberg correction (two-sided one-sample Wilcoxon
signed-rank tests, $q<10^{-14}$). Resilient networks showed smaller
decreases in $\bar r_E$ ($-29.90\%$), $\bar r_I$ ($-28.00\%$), and
$\bar h_{E\to E}$ ($-24.81\%$; all $q<10^{-14}$), while
$|\bar h_{I\to E}|$ increased by $5.38\%$ ($q<10^{-12}$).
The stress-induced change differed between naive and resilient networks for all four
quantities (two-sided Mann--Whitney tests, $U=0$ and
Benjamini--Hochberg-adjusted $q<10^{-17}$ in each comparison). The same
between-class ordering for all four quantities was present at weaker stress with $\delta=0.25$ (all adjusted
$q<10^{-17}$ for between-class comparisons). Thus, only resilient networks showed genuine inhibitory dominance:
$|\bar h_{I\to E}|$ increased in absolute terms, whereas in naive networks it
fell alongside excitation, so the drop in $E/|I|$ there reflected a faster
loss of excitatory drive rather than a strengthening of inhibition. 
As reported above, this
selective preservation of inhibitory drive in resilient networks was a
population-level property, not a peculiarity of the example network.

Resilience training also changed and stabilised the network's energetic regime
(Fig.~\ref{fig:figure2}e). Here, metabolic energy is a per-timestep proxy that
combines the costs associated with recurrent population activity and synaptic
transmission, calculated using the effective stress-perturbed weights and
averaged across trials and time (Methods,
Section~\ref{sec:methods_energy_analysis}). Without applied stress,
resilient networks operated at a lower energy level than naive networks
($0.900$ versus $1.066$; $U=1938$, $p=2.14\times10^{-6}$, Fig.~\ref{fig:figure2}e, BL). Stress then
produced a pronounced energy-consumption collapse in naive networks, while resilient networks remained much closer to their own lower baseline energy, see Fig.~\ref{fig:figure2}e, $\delta\in\{0.25,0.5\}$. Relative to each
network's own baseline, these changes corresponded to mean energy deviations of
$39.8\%$ versus $6.8\%$ at $\delta=0.25$ and $55.0\%$ versus $10.4\%$ at
$\delta=0.5$ (naive versus resilient; both $U=2500$, $p<10^{-17}$). 
In conclusion, resilience training maintains task performance under stress by conserving activity trajectories, which also implies stable energy measures. Without stress, energy consumption is reduced as compared to naive networks, likely due to an overall decrease in activity and synaptic drive (Fig.~\ref{fig:figure2}b). Overall, resilience training stabilises network activity, inhibition and energy across stress levels, at the cost of decreased task generalisation investigated in detail in the next section.

\subsection{The resilience--generalisation trade-off persists across training conditions}
\label{sec:delay-tradeoff-analysis}

\begin{figure}[ht!]
    \centering
    \includegraphics[width=0.98\linewidth]{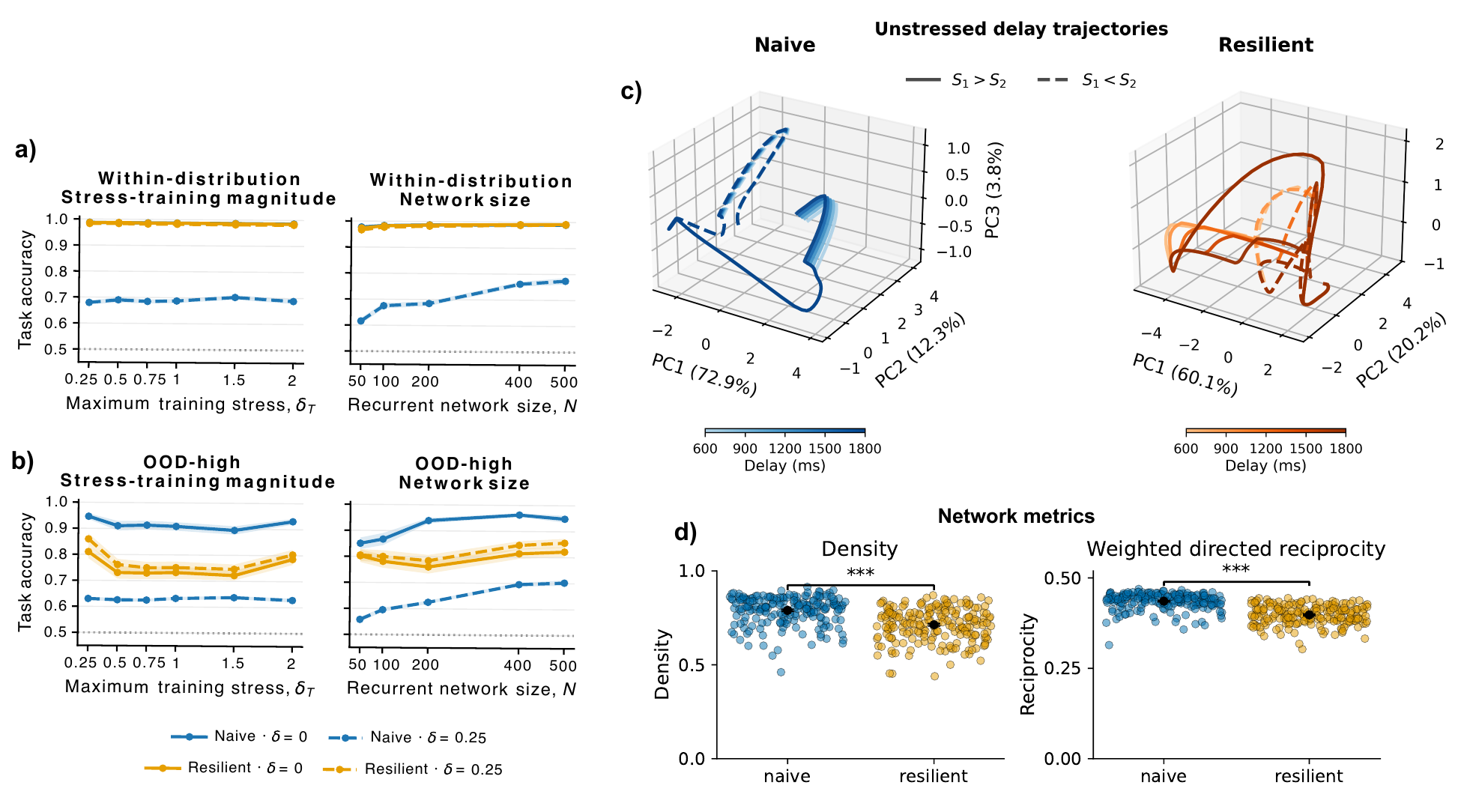}
    \caption{%
    \textbf{The resilience--generalisation trade-off is robust to training
    stress magnitude and network size and is accompanied by distinct dynamics and
    network topology.}
    \textbf{(a,b)} Accuracy as a function
    of the maximum resilience-training stress $\delta_T$ (left) or recurrent
    network size $N$ (right) within the trained delay distribution (a) or long out-of-distribution delays (OOD-high, b) for the
    same sweeps. Blue and orange denote naive and resilient networks evaluated without stress ($\delta=0$, solid) and with applied stress ($\delta=0.25$, 
    dashed). Points are means and bands are SEMs ($n=50$ networks per
    class and sweep point).
    \textbf{(c)} Unstressed recurrent trajectories for the representative naive
    and resilient networks from Fig.~\ref{fig:figure2} across delays from
    $600$ to $1800\,\mathrm{ms}$. Darker colours encode longer delays and line style
    encodes choice.
    \textbf{(d)} Recurrent-network density and weighted directed reciprocity.
    Points show trained networks ($n=200$ per class) and black markers show
    means$\pm$SEM; *** $p<0.001$ by two-sided Mann--Whitney \textit{U} tests, $U=30414$ for density, $U=35182$ for reciprocity.}
    \label{fig:figure3}
\end{figure}

We tested whether the trade-off could be compensated by varying the maximum stress
magnitude used during resilience training, $\delta_T$, or by increasing the
number of recurrent neurons, $N$ (Fig.~\ref{fig:figure3}a,b). Each sweep point
contained 50 independently trained networks per class and was evaluated on
$2000$ trials per exact delay, both without stress and with applied stress of $\delta=0.25$. Across the complete analysis, this comprised 11 sweep points
and 1100 networks. Welch $t$-tests compared network classes, and ordinary least-squares linear models with heteroskedasticity-consistent type 3 (HC3) standard errors tested network-class-by-sweep interactions.

A resilience-generalisation trade-off was observed across various levels of training stress and network sizes
(Fig.~\ref{fig:figure3}a,b), with the naive network performing better than the resilient network for long delays without stress. Network performance showed no significant interaction with training stress magnitude (interaction $p=0.208$ for within-distribution, $p=0.881$ and $p=0.206$ for out-of-distribution).
Full group-comparison and interaction statistics are reported in Supplementary
Table~\hyperref[tab:delay_tradeoff_stats]{S4}.

Increasing network size partially rescued naive-network performance under
applied stress, and this benefit was significantly greater for naive than
resilient networks (Fig.~\ref{fig:figure3}a,b, right). Within the trained delay
range, stressed naive accuracy increased by $0.033$ per additional $100$
recurrent units (HC3 $95\%$ CI $[0.027,0.039]$,
$p=2\times10^{-28}$).
The naive slope exceeded the resilient slope by $0.028$ per $100$ units
($95\%$ CI $[0.022,0.034]$, interaction $p=2\times10^{-21}$), narrowing
the stressed accuracy gap between resilient and naive network from $0.35$ to $0.22$ without eliminating it.
Averaged across sizes, within-distribution stressed accuracy remained $0.70$
in naive and $0.98$ in resilient networks ($p=3\times10^{-126}$), for all tested network sizes, resilient networks show better task performance than naive networks under stress.
The same partial rescue was evident at OOD-high delays: stressed naive accuracy
increased by $0.032$ per $100$ units ($95\%$ CI $[0.027,0.037]$,
$p=5\times10^{-41}$), 
and the naive slope exceeded the resilient slope by $0.018$ per $100$ units
($95\%$ CI $[0.007,0.029]$, interaction $p=0.0013$). Nevertheless, under stress,
resilient networks perform better than naive networks even under OOD-high delays (Fig.~\ref{fig:figure3}b, right). 
However, without test-time
stress, naive networks outperform resilient networks across network sizes, with the advantage slightly decreasing with network size (interaction $p=0.036$).

To determine whether this behavioural generalisation cost had a corresponding
population-dynamical signature, we extended the PCA trajectory analysis used
above for trajectories under stress in Fig.~\ref{fig:figure2}a. On unstressed baseline networks, we 
compared trajectories at delays of
$600$, $900$, $1200$, $1500$, and $1800\,\mathrm{ms}$. For visualisation, the
mean trajectories of representative naive and resilient networks were
projected into the PCA space of the 
unstressed baseline activity (Fig.~\ref{fig:figure3}c), the selected networks were the 
same representative networks from Section~\ref{sec:recurrent_dynamics_geometry}. 

For the quantitative
analysis across all 50 networks per class, trajectories were evaluated
in each network's fixed baseline PCA subspace. End-of-delay displacement was
defined as the choice-averaged Euclidean distance between the mean recurrent
state during the final $100\,\mathrm{ms}$ before the second stimulus and the
corresponding state at the largest trained delay, $900\,\mathrm{ms}$, serving here as reference delay.
Network classes were compared at each delay using two-sided Mann--Whitney $U$
tests. Holm correction was applied jointly across nine planned delay-geometry
comparisons: four end-of-delay displacement comparisons and five
choice-separation comparisons.
Relative to the $900\,\mathrm{ms}$ reference, end-of-delay displacement became
larger in resilient networks for all delays longer than the trained regime: At
$1200\,\mathrm{ms}$ it was $0.69\pm0.12$ in naive and $1.33\pm0.14$ in
resilient networks (Holm-adjusted $p=0.016$); at $1500\,\mathrm{ms}$,
$1.17\pm0.18$ versus $2.04\pm0.19$ ($p=0.014$); and at
$1800\,\mathrm{ms}$, $1.33\pm0.18$ versus $2.18\pm0.20$ ($p=0.013$).
At a trained delay of $600\,\mathrm{ms}$, the difference did not survive correction ($p=0.20$).

The representative projections therefore illustrate a population-level
difference in how the trained geometry is extrapolated: naive trajectories
departed more gradually from the $900\,\mathrm{ms}$ representation, whereas
resilient trajectories underwent a larger delay-dependent displacement. Together,
these results are consistent with naive networks retaining a more controlled
representation for extrapolation to longer delays, while resilient networks
follow a more delay-specific dynamical solution outside the trained regime.

Resilience training was also associated with a change in recurrent topology
(Fig.~\ref{fig:figure3}d). Network density decreased from
$0.79\pm0.01$ in naive networks to $0.71\pm0.01$ in resilient networks
($U=30414$, $p=2\times10^{-19}$). Weighted directed reciprocity likewise
decreased from $0.44\pm0.002$ to $0.40\pm0.002$
($U=35182$, $p=2\times10^{-39}$). The resilient solutions were therefore
less dense and less reciprocally coupled, consistent with a more selective
recurrent organisation that stabilised the trained computation while offering
less flexibility for extrapolation.

\begin{figure}[t!]
    \centering
    \includegraphics[width=\linewidth]{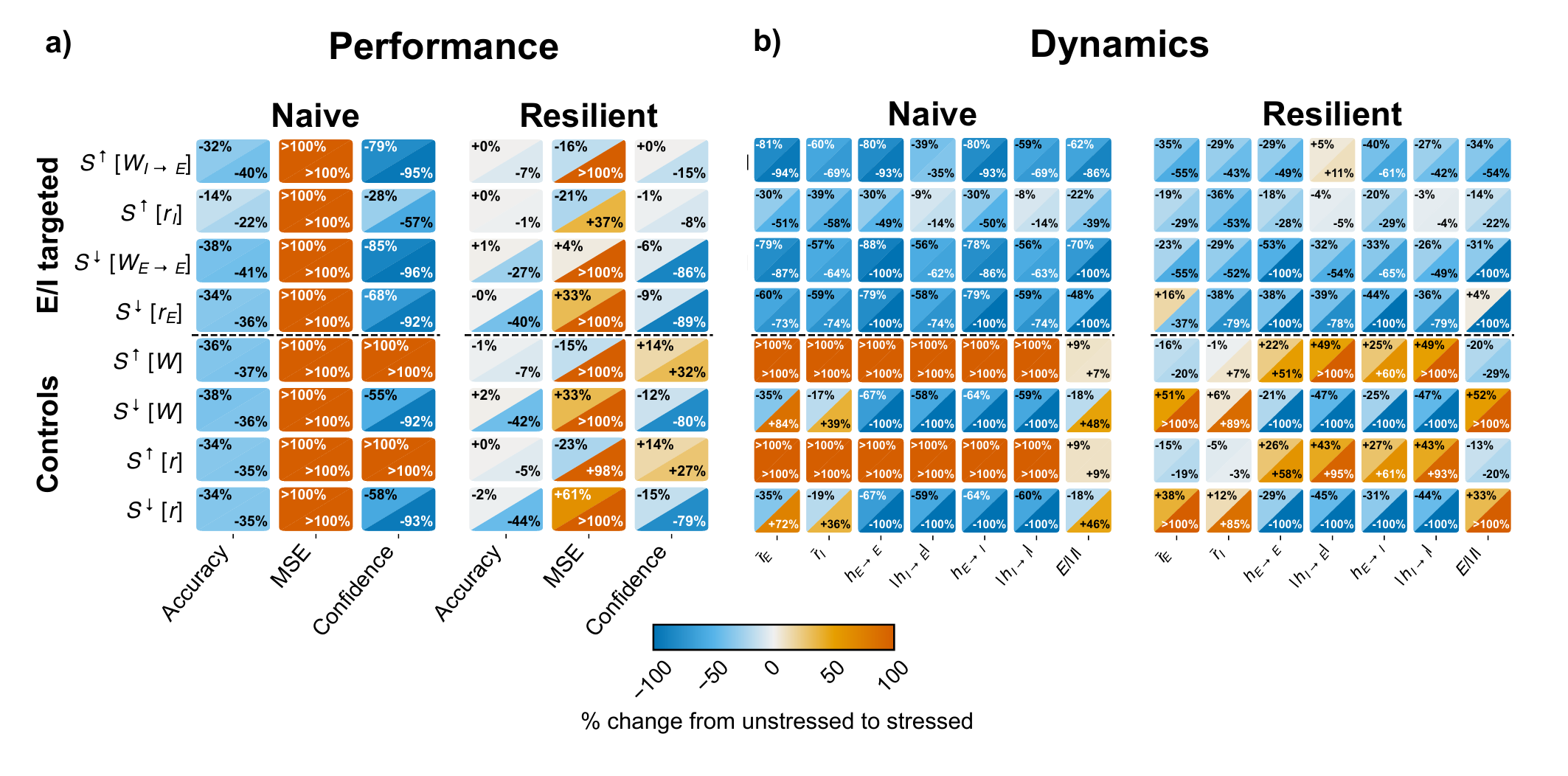}
    \caption{%
    \textbf{Comparison of targeted and non-selective putative stress operators.}
    Each cell shows the median signed percentage change from the unstressed
    baseline across 50 independently trained networks. The upper-left triangle
    reports the effect of $\delta_T=0.5$ and the lower-right triangle $\delta_{\max}=1.0$;
    blue denotes a decrease and orange an increase, with the colour scale
    saturated at $\pm100\%$. Rows above the dashed line are targeted E/I
    perturbations and rows below it are population-agnostic controls; arrows
    indicate up- or down-scaling of the bracketed weights or activities.
    Naive networks were trained without perturbation, whereas each resilient
    group was trained with the operator shown in its row. \textbf{(a)} Changes
    in accuracy, readout mean-squared error (MSE), and decision confidence.
    \textbf{(b)} Changes in excitatory and inhibitory population activity
    ($\bar r_E$, $\bar r_I$), recurrent source-to-target drives
    ($h_{E\to E}$, $|h_{I\to E}|$, $h_{E\to I}$, $|h_{I\to I}|$), and the
    excitatory-to-inhibitory drive ratio $E/|I|$. $\SEI$ in the first row is
    the only targeted operator that combines preserved resilient accuracy with
    excitatory hypofunction, a lower $E/|I|$ ratio, and maintained or increased
    inhibitory drive onto excitatory units.}
    \label{fig:stress_operator_comparison}
\end{figure}

\subsection{Comparison with alternative stress operators}
\label{sec:stres_mech_comparison}

So far, we have analysed the stress operator $\SEI$ that increases inhibitory synaptic strength. We here show that the effects of $\SEI$ fit experimental observation on chronic stress better as compared to alternative putative stress operators. Comparing simulation results to the experimentally motivated signatures from Table~\ref{tab:targets}, we analysed selective changes to recurrent
excitation or population gain with non-selective up- and down-scaling controls. Each heat-map cell of Fig.~\ref{fig:stress_operator_comparison} compares an
operator with the same network's unstressed baseline; its two triangles report the changes with respect to unstressed baseline for 
the maximum training magnitude, $\delta_T=0.5$ (upper triangle), and a stronger out-of-range
test, $\delta_{\max}=1.0$ (lower triangle).

The naive networks show responses to applied stress that alone did not clearly distinguish between targeted mechanisms:
several operators lowered $E/|I|$ by suppressing both excitatory and inhibitory
drive. The resilient networks showed distinct patterns for different putative stress operators. At $\delta_T=0.5$, $\SEI$
increased $|h_{I\to E}|$ by $5.4\%$ while reducing $h_{E\to E}$ by $29.4\%$,
$\bar r_E$ by $34.7\%$, and $E/|I|$ by $33.9\%$ (all BH-adjusted
$q<10^{-10}$), with no detectable change in accuracy ($+0.1\%$, $q=0.89$).
The same pattern compatible with experimental observations (Table~\ref{tab:targets}) strengthened at $\delta_{\max}$ for $\SEI$. Alternative targeted
operators either reduced inhibitory drive together with excitation or produced
a weaker E/I shift, and the non-selective controls generated qualitatively
different responses. Thus, this secondary comparison supports $\SEI$ as the
canonical operator used above; full operator definitions and analysis details
are provided in the Supplemental Information, S2 Text.

\section{Discussion}

Using Dale-constrained recurrent networks, we asked how a biologically motivated circuit-level
stress perturbation changes working-memory computation and how adaptation to
that perturbation reshapes performance under stress and task generalisation. Together, our results show that adaptation to selective stochastic upscaling of inhibitory synapses onto excitatory units ($\SEI$) is sufficient to produce the chronic-stress-like phenotype examined here. Whereas naive networks trained without stress showed large losses in decision confidence and accuracy when the stress perturbation was reapplied at test time, networks trained in the presence of $\SEI$ maintained psychometric and delay-task performance under applied stress. This functional resilience was accompanied by a shift in recurrent drive towards inhibitory dominance, more constrained population dynamics, a more stable energetic measure, altered recurrent topology compatible with increased feed-forward processing, and reduced generalisation to untrained long delays.

Under $\SEI$, naive networks underwent a broad loss of recurrent activity: under stress-increased inhibition, the magnitude of both excitatory and inhibitory drive fell due to population interactions. The observed reduction in the $E/|I|$ ratio thus did not reflect strengthened inhibitory signalling, but resulted from excitation decreasing faster than inhibition. Resilient networks also show a decrease in excitatory drive (and activity) under stress, but in contrast to naive networks, resilient networks maintain or increase inhibitory drive onto excitatory
neurons. Therefore, genuine inhibitory dominance emerged as an {\it adapted circuit
state}, not as the immediate consequence of upscaling inhibitory weights. This
asymmetric response shows that a single mechanism, increased inhibition under chronic stress, is sufficient to explain distinct experimental observations of elevated
PV-interneuron activity~\cite{Page2019-xg}, increased GABAergic contacts
onto pyramidal cells~\cite{McKlveen2016-px}, and concurrent increases in inhibition
and decreases in excitation across prefrontal subregions~\cite{Rodrigues2024-ry}.

The comparison with alternative operators sharpens our interpretation of
$\SEI$ as reasonable stress operator. Several perturbations lowered the excitatory-to-inhibitory-drive ratio by
suppressing excitatory and inhibitory drive together, which is not in agreement with the biological signatures from Table~\ref{tab:targets}. Among the targeted operators, $\SEI$  uniquely combined preserved accuracy with excitatory hypofunction, a lower $E/|I|$ ratio, and
maintained or increased absolute inhibitory drive onto excitatory units. 
Thus, among the operators tested, $\SEI$ is the most coherent perturbation on synaptic weights and neuronal excitability for reproducing the joint chronic-stress signatures from Table~\ref{tab:targets}, which motivated us to study the networks' adaptation to this stress operator with the aim to identify fundamental network strategies that allow for stress resilience.

The population and energetic analyses converge on how resilience was
implemented. Under applied $\SEI$, resilient trajectories remained closer to their unstressed baseline than naive trajectories and largely within the same baseline PCA subspace. Thus, resilience reflects restricted movement within the existing population geometry rather than recruitment of a distinct subspace.
For the energetic analysis we used a model proxy of biological metabolic expenditure, basically a sum over neuronal activity and synaptic drive. Naive networks showed substantial reductions in activity- and synapse-related energy under stress, whereas resilient networks operated from a lower-energy baseline that remained comparatively stable under perturbation. Since
energetic stability was not an optimisation target, its emergence alongside
dynamical stability suggests that these are linked features of the learned robust
solution. 
In agreement with an overall reduced activity level of resilient networks, resilient connectivity was less dense and less reciprocally coupled than 
naive connectivity, possibly resilience is mediated partly by a more local and thus restricted action of stress-strengthened inhibition.

A central finding of this work is that resilience carries a generalisation cost, 
and this trade-off persists across training conditions.
Resilient networks matched naive performance within the training distribution
when unstressed, and strongly outperformed them under $\SEI$, whereas naive
networks generalised better to unstressed long delays between the two stimuli, a manipulation that challenges the working-memory of the trained network. Possibly, stronger feed-forward-oriented processing of resilient networks reduces network capacity for working-memory, which mostly depends on recurrent activity. 
Better generalisation of the naive network without stress
persisted across the full sweep of resilience-training stress magnitudes,
indicating the trade-off does not depend on a particular choice of training magnitude.
Increasing network size
partially rescued naive performance under stress, more so than for resilient
networks. Changing network size did however neither eliminated the better performance of resilient networks under applied stressed, nor did it 
reverse the better performance of the naive networks under long-delay stimuli. Capacity can therefore buffer
vulnerability to $\SEI$, but does not by itself explain or resolve the reported trade-off.

The resilience-generalisation trade-off is reflected by the population dynamics observed without stress. Under unstressed baseline conditions, the recurrent population trajectory of resilient networks diverged strongly from their trained $900\,\mathrm{ms}$ reference trajectory as stimuli delays extended into the untrained, out-of-distribution range, whereas the trajectories of naive networks changed less and more gradually.

Our framing suggests a more specific interpretation of stress-related
inflexibility. Resilience does not mean that a circuit is unchanged by stress.
Here resilience emerged through network reorganisation into a stable, lower-energy operating
regime that is able to maintain a familiar computation under a familiar perturbation.
The cost appeared when the same computation involving working-memory had to be extrapolated beyond its
trained temporal regime. Habit-like behaviour reported after chronic stress~\cite{Dias-Ferreira2009-nh,Girotti2024-vt} may therefore reflect, in part, the
specialisation of a stress-adapted circuit rather than a uniform loss of computational
ability. A direct experimental prediction of our results is that a stress-adapted prefrontal
circuit can preserve performance under elevated inhibitory drive, introduced for example by optogenetic manipulations, while the corresponding 
population activity should become increasingly displaced when the working-memory interval is
extended beyond familiar durations, for example using mice trained on a working-memory task.

More broadly, the study illustrates the value of training biologically
constrained recurrent networks~\cite{Song2016-qj,Yang2019-sq} to analyse resilience and stress mechanisms. Dale's principle and explicit excitatory and
inhibitory populations allow perturbations to be defined at the level of a
specific recurrent projection, while complete access to activity, synaptic
drive, geometry, connectivity, and energetic proxies connects the stress perturbation
to computation. Modelling stress as an external operator on specific neuronal parameters isolates its circuit-level effects from the endocrine, immune, and behavioural processes that normally shape the complex biological stress response \cite{Godoy2018-js,McEwen2016-wi}. This abstraction facilitates mechanistic analysis to understand the specific contributions of our stress operator while also defining the scope of our conclusions.

\paragraph{Limitations.}
The main limitation is that $\SEI$ is an imposed multiplicative perturbation,
not a model of how chronic stress develops biologically. The operator comparison
was restricted to a finite set of isolated perturbations and therefore supports
$\SEI$ as the best match within that set, not as a unique biological mechanism.
Real stress responses coordinate molecular, endocrine, neuromodulatory, and
circuit processes across timescales~\cite{Joels2009-gm}, and the present model
does not describe exposure history, plasticity during stress, recovery, or the
transient dynamics of acute stress. The rate network also uses a strict form of
Dale's principle and omits interneuron subclasses, conductance dynamics and
co-transmission~\cite{Svensson2018-xd}. As in all trained
RNN models, architecture, objective, optimisation, and task statistics can
change the learned solution~\cite{Sussillo2013-fg,Dubreuil2021-sf,Kao2019-ns},
and the networks carry no experiential history beyond their training~\cite{Ma2020-to}. Finally, the generalisation trade-off was shown for one temporal
extrapolation in one categorical working-memory task. 

\paragraph{Future directions.}
Several extensions follow directly. First, $\SEI$ could be generated by learned
cell-type-specific plasticity rather than imposed externally, closing the loop
between stress exposure and circuit adaptation. Combining $\SEI$ with
neuromodulatory gain, or excitatory perturbations would test whether distinct
physiological routes converge on the same robust phenotype. Second, targeted
interventions on density, reciprocal motifs, or the synapses that contribute
most strongly to trajectory stability could determine whether the observed
topology mediates robustness and/or the delay-generalisation trade-off. Third, training
on continuous-attractor dependent tasks~\cite{Driscoll2024-rd}, multitask demands~\cite{Yang2019-sq}
or naturalistic task statistics could reveal whether the delay-generalisation
trade-off is a general property of stress adaptation or specific to categorical
comparison tasks.
Finally, and perhaps most promising, modelling stress dynamics within a trial would connect the present
chronic-stress account to the transient dynamics of acute stressor exposure. This would 
require networks capable of online adaptation, and would establish how far the
present account extends toward biological stress resilience and stress inoculation.

\section{Methods}

\subsection{Network model}\label{sec:network_model}
We follow the formulation of Song \textit{et al.}~\cite{Song2016-qj}, using a rate-based recurrent
neural network with excitatory and inhibitory populations. Unless stated
otherwise, networks comprised N=200 neurons, of which $N_E = 0.8 N$ were excitatory and
$N_I = 0.2 N$ inhibitory. Before training, the networks are initialised as fully connected with weight matrix $W$, with self-connections removed. Each neuron is described by an internal state (a membrane-potential analogue),$x^i_P(t)$, combined into two vectors $x_P(t)$, where $P \in \{E, I\}$ denotes the population. The activity profile of each population is represented by the instantaneous firing rate, given by the population activity vectors $r_P(t) = \phi(x_P(t))$, where $\phi(\cdot)$ is the rectified linear unit (ReLU), $\phi(x) = \max(0, x)$. The network evolves in discrete time by a first-order update, with the state of a neuron in population $P$ at time $t+ \Delta t$ given by

\begin{equation} 
x_P(t+ \Delta t) = (1-\alpha)\,x_P(t) + \alpha\,h_P(t) + \alpha\,s_P(t) + \sqrt{2\alpha}\,\sigma_\xi\,\xi_P(t), 
\end{equation}

where the vector $h_P(t)$ is the recurrent synaptic drive defined below, $s_P(t) = W_{\mathrm{in},P}\, u(t)$ is the input drive to population $P$, $\alpha = \Delta t / \tau_r$ sets the integration timescale, and $\xi_P(t) \sim \mathcal{N}(0, I)$ is an independent Gaussian noise of amplitude $\sigma_\xi = 0.02$. Simulations used a time step $\Delta t = 20$~ms and a membrane time constant $\tau_r = 100$~ms.

The recurrent synaptic drive collects excitatory and inhibitory contributions,
\begin{align} 
h_E(t) &= \underbrace{W_{EE}\,r_E(t)}_{h_{E\to E}(t)} 
        + \underbrace{W_{EI}\,r_I(t)}_{h_{I\to E}(t)}, 
        \label{eq:main_drives_E} \\
h_I(t) &= \underbrace{W_{IE}\,r_E(t)}_{h_{E\to I}(t)} 
        + \underbrace{W_{II}\,r_I(t)}_{h_{I\to I}(t)}, 
        \label{eq:main_drives_I} 
\end{align}

where $r_E(t)$ and $r_I(t)$ are the population activity vectors defined above and 
$W_{PQ}$ is the recurrent connectivity from population $Q$ to population $P$. While the input drive projects the task to both populations, the output reads only from the excitatory population, $y(t) = W_{\mathrm{out}}\,r_E(t)$. This mirrors the organisation of cortical circuits: thalamic afferents target both excitatory and inhibitory neurons, whereas long-range output to downstream areas is carried predominantly by excitatory pyramidal cells, with inhibitory interneurons acting mainly within local circuitry~\cite{Harris2015-gh}.

Following Song \textit{et al.}~\cite{Song2016-qj}, recurrent weights were initialised as positive magnitudes and assigned excitatory or inhibitory signs according to Dale’s principle~\cite{Dale1935-yk,Strata1999-fv,Svensson2018-xd}, \textit{i.e.} positive sign for $W_{EE}$ and $W_{IE}$, and negative sign for $W_{EI}$ and $W_{II}$. Using fully connected recurrent networks, excitatory and inhibitory presynaptic weights were sampled from Gamma distributions ($k=2$) with pre-rescaling means $\mu_E=1$ and $\mu_I=N_E/N_I$, respectively. Self-connections were removed, and the signed recurrent matrix, denoted as $W_{\mathrm{rec}}$, 
was rescaled to a spectral radius of $1.5$. Input and output weights were drawn from rectified Gaussian distributions with scale $1/\sqrt{N}$. During training, recurrent magnitudes and input and output weights were constrained to be non-negative through a ReLU parameterisation, thereby preserving Dale's principle. The initial states $x_0$ were initialised as $x^i_0\sim\mathcal N(0,0.1^2)$, and optimised jointly with the synaptic parameters.

\subsection{Delayed parametric comparison task implementation}\label{sec:task_details}

The delayed parametric comparison task adapts the parametric working-memory paradigm in which prefrontal delay activity encodes remembered stimulus values~\cite{Romo1999-la,Barak2010-fk}. Each network received three input channels and produced two output channels. The input channels encoded fixation, the first scalar stimulus $S_1$, and the second scalar stimulus $S_2$, respectively. The two output channels encoded the choices $S_1>S_2$ and
$S_2>S_1$. Trials used a time step of $\Delta t=20$ ms for forward integration.

Each trial lasted 2000 ms, corresponding to $T=100$ simulation steps. Trials
began with a 400 ms fixation epoch. Then, each stimulus was presented for
250 ms with the delay between $S_1$ and $S_2$
sampled uniformly from 400--900 ms. The decision epoch began immediately after the second stimulus and continued until the end of the trial, as illustrated in Fig.~\ref{fig:figure1}.

Stimulus amplitudes were restricted to the interval from 0.2 to 1.2, $S_1, S_2 \in [0.2, 1.2]$. Trial difficulty was controlled by the signed difference $\Delta = S_2 - S_1$. For training, difference magnitudes were sampled from $|\Delta| \in \{0, 0.032, 0.064, 0.128, 0.256, 0.512\}$. For non-zero differences, the sign was chosen randomly, so trials included both
$S_1>S_2$ and $S_2>S_1$ comparisons. For a given sampled $\Delta$, $S_1$ was
drawn uniformly from the range that allowed $S_2=S_1+\Delta$ to remain within
the stimulus bounds. The fixation channel was active from trial onset until the decision epoch. A tonic baseline input of $0.2$ was added to all input channels throughout the trial, and input noise was added independently across time points and input channels, sampled log-uniformly from $\sigma_{\mathrm{in}} \in [0.01, 0.05]$ and rectified to be non-negative.

Targets are zero during the trial time span except during the decision period. If $S_1>S_2$, the target was $[1,0]$; if $S_2>S_1$, the target was $[0,1]$. For exact ties
($S_1=S_2$), the target was $[0.5,0.5]$. The training mask assigned full loss
weight of $m = 1$ during fixation and decision periods. The intervening stimulus and delay
periods were assigned a reduced mask weight of $m = 0.01$, so that training emphasised fixation maintenance and final choice accuracy while
only weakly constraining network output during stimulus presentation and memory
maintenance.

\subsection{Training objective}

Networks were trained by minimising a temporally weighted mean-squared error
between the readout of minibatch $b$ at timestep $t$, $y_{bt}=W_{\mathrm{out}}r_{E,bt}$ and target
$\hat y_{bt}$, together with a firing-rate regularisation term:
\begin{equation}
\mathcal{L}(\theta)
=
\frac{
\sum_{b=1}^{B}\sum_{t=1}^{T}
m_{bt}\left\lVert y_{bt}(\theta)-\hat y_{bt}\right\rVert_2^2
}{
\sum_{b=1}^{B}\sum_{t=1}^{T}m_{bt}+\epsilon
}
+
\lambda_r\frac{1}{BTN}
\sum_{b=1}^{B}\sum_{t=1}^{T}\sum_{i=1}^{N}
r_{bti}(\theta)^{2},
\end{equation}
where $\theta$ denotes model parameters, $B$ is the minibatch size, $T$ is the number of simulation steps, and
$N$ is the number of recurrent units. We used $\epsilon=10^{-6}$ to ensure a denominator larger than zero for numerical
stability and $\lambda_r=10^{-3}$ for firing-rate regularisation. The trainable
parameters $\theta$ comprised the parameters underlying
$W_{\mathrm{rec}}$, $W_{\mathrm{in}}$, and $W_{\mathrm{out}}$, together with
the initial state $x_0$. The temporal mask $m_{bt}$ was set to $m_{bt}=1$ during the
initial fixation and decision epochs and to $m_{bt}=0.01$ during stimulus presentation
and the intervening delay. This weighting emphasised maintaining zero output
during initial fixation and producing the correct choice during the decision
period, while only weakly constraining output to zero during stimulus presentation and
memory maintenance. Networks were optimised using Adam with a learning rate of $10^{-3}$ and
minibatches of $B=128$ independently generated trials. The global gradient norm
was clipped at $1$. Networks used in the analyses reported in Results
Sections~\ref{sec:results_performance} and
\ref{sec:recurrent_dynamics_geometry} were trained for $8000$ optimisation
steps.

\subsection{Stress mechanisms}\label{sec:stress_mecs}
We model chronic stress as a random operator $S$ acting on a target $X$, either a
recurrent weight submatrix $W_P$ or a population activity vector $r_P$, by multiplying it
element-wise by a random factor $g$,
\begin{equation}
  S[X] = g \odot X,
  \qquad
  g_k = \exp(\eta + \tau_g Z_k),
  \quad Z_k \sim \mathcal{N}(0,1)\ \text{i.i.d.},
  \label{eq:stress_operator}
\end{equation}
drawing one factor $g_k$ independently for each weight or neuron in the target. The
magnitude $\delta > -1$ and relative dispersion $\sigma_{\mathrm{stress}} \geq 0$
(fixed at $\sigma_{\mathrm{stress}} = 0.2$) enter through
\begin{equation}
  \tau_g = \sqrt{\log\!\big(1+\sigma_{\mathrm{stress}}^2\big)},
  \qquad
  \eta = \log(1+\delta) - \tfrac{1}{2}\tau_g^2 .
  \label{eq:stress_params}
\end{equation}
The $-\tfrac{1}{2}\tau_g^2$ term centres the factors so that, on average, $S$ rescales its
target by $(1+\delta)$. Since $\mathbb{E}[\exp(\eta+\tau_g Z)] = \exp(\eta+\tfrac{1}{2}\tau_g^2)$
for $Z\sim\mathcal{N}(0,1)$,
\begin{equation}
  \mathbb{E}[g_k] = 1+\delta,
  \qquad
  \mathrm{CV}[g_k] = \sqrt{\exp({\tau_g^2})-1} = \sigma_{\mathrm{stress}},
  \label{eq:stress_moments}
\end{equation}
so $\delta$ sets the average change and $\sigma_{\mathrm{stress}}$ its spread, separately,
and
\begin{equation}
  \mathbb{E}\big[S[X]\big] = (1+\delta)\,X .
  \label{eq:stress_expectation}
\end{equation}
We write $S^{\uparrow}$ for upscaling ($\delta>0$) and $S^{\downarrow}$ for the matched
operator with $\delta$ negated ($\delta<0$), and state the magnitude where needed (for example
$S^{\uparrow}[W_{EI}]$ with $\delta = 0.25$). The baseline network is recovered at
$\delta=0$, $\sigma_{\mathrm{stress}}=0$, where $g\equiv 1$ and $S[X]=X$.

\paragraph{Synaptic weight scaling.}
For the canonical operator $\SItoE$, independent factors from
Eq.~\ref{eq:stress_operator} multiply the elements of $W_{EI}$, giving
$\mathbb{E}[\widetilde W_{EI}]=(1+\delta)W_{EI}$. The alternative weight and
activity-gain operators used only in the comparison are detailed in the Supporting
Information, S2 Text.

\subsection{Resilience training}\label{sec:stress_training}
We compare two training regimes. Under standard training, no stress operator is
applied. Under resilience training, each trial draws its own stress instance, with
magnitude $|\delta| \sim \mathcal{U}[0,\delta_T]$ and sign determined by the
operator direction. The sampled factors are held fixed throughout that trial's
forward pass. The stress operator $S$ is applied to its target according to
the mechanism under study, perturbing either the effective weights derived from
$\theta$ or the population activity $r(t)$, and the task loss is computed from
the output of the resulting perturbed forward pass. Because the perturbation is
an element-wise product by sampled factors that are held fixed throughout the
trial, it is differentiable with respect to the underlying parameters  $\theta$, and
gradients flow back to $\theta$ whether the operator acts on weights or
activity. Drawing a new stress instance for each trial trains the network to
solve the task across perturbations with magnitudes up to $\delta_T$, rather
than under a single unperturbed configuration.

For the primary analyses, the naive and resilient groups each consisted of 50
independently trained networks, with $\delta_T=0.5$ for the resilient group.
The training-stress-magnitude analysis used separately trained resilient groups
at each value of $\delta_T$. For the secondary operator comparison
(Section~\ref{sec:stres_mech_comparison}), a separate resilient group was
trained for each alternative operator and used only in that comparison.

\subsection{Systematic stress mechanism comparison}\label{sec:methods_stress_evaluation}
We evaluated eight candidate operators, see Fig.~\ref{fig:stress_operator_comparison} and Supplemental Information, in naive networks and networks trained
for resilience to the corresponding operator. Fig.~\ref{fig:stress_operator_comparison}
reports the median signed change from each network's unstressed baseline at
$\delta_T=0.5$ and $\delta_{\max}=1.0$ across 50 independently trained networks. 
The complete operator definitions, sampling scheme, performance and dynamics
metrics, aggregation procedure, and statistical tests are given in the Supporting Information,
S2 Text.

\subsection{Recurrent activity and drive analysis}
\label{sec:methods_recurrent_drives}

For the canonical $\SItoE$ analysis, each trained network was evaluated on the
same fixed set of 5000 task trials at $\delta\in\{0,0.25,0.5\}$. A separate
stress instance was sampled for each stressed trial. Population activities
$\bar r_E$ and $\bar r_I$ and recurrent drives $\bar h_{E\to E}$ and
$|\bar h_{I\to E}|$ were averaged across all trial time points, yielding one
value per network and stress condition. For each nonzero stress magnitude, we
expressed each network's value as a signed percentage change from its own
unstressed baseline,
\begin{equation}
\Delta\%_{m,s}(\delta)=100\times
\frac{m_s(\delta)-m_s(0)}
{\left|m_s(0)\right|},
\label{eq:pct_change}
\end{equation}
where $m_s$ denotes an activity or drive measure of an independently
trained network $s$. The network was the unit of statistical replication ($n=50$
per class).

Within each network class, percentage changes were tested against zero using
two-sided one-sample Wilcoxon signed-rank tests. Because naive and resilient
networks were trained independently, their percentage changes were compared
using two-sided Mann--Whitney $U$ tests; the corresponding test statistics and
adjusted $q$ values are reported in
Section~\ref{sec:recurrent_dynamics_geometry}.
Matching numerical seed identifiers
across classes did not constitute statistical pairing. Benjamini--Hochberg
false-discovery-rate correction was applied
separately to the 16 planned within-class tests (four measures, two network
classes, and two stress levels) and the eight planned between-class tests (four
measures and two stress levels). Adjusted values are reported as $q$ values.

\subsection{Recurrent activity PCA analysis}\label{sec:methods_pca_analysis}

For each network, PCA was fitted to the baseline population activity collected from a 
fixed set of 5000 trials. The first three baseline PCs were used only for 
visualisation in Figure~\ref{fig:figure2}. Activity from each stress condition 
was then projected onto the same baseline PCs, allowing baseline and stressed 
trajectories to be compared in a common coordinate system within each network.

We then selected the smallest number of PCs needed to explain at least $90\%$ 
of baseline variance,

\begin{equation}
k =\min \left\{k'\in \{1,\ldots,N\} :\frac{\sum_{i=1}^{k'} \lambda_i}{\sum_{i=1}^{N} \lambda_i}\geq 0.90\right\},
\end{equation}

where $\lambda_i$ are the ordered baseline PCA eigenvalues, tr denotes the trace. These $k$ PCs defined the baseline subspace and were kept fixed for all stress conditions. For each stress level $\delta$, activity was centred using the baseline mean, $\tilde X^{\delta} = X^{\delta} - \mu$, and its covariance $C^\delta$ was computed. The subspace preservation index was defined as the fraction of stressed-condition variance captured by the baseline subspace:

\begin{equation}
\rho(\delta)=\frac{\operatorname{tr}\left(P\, C^\delta \,P^\top\right)}{\operatorname{tr}\left(C^\delta\right)} ,
\end{equation}

where $P \in \mathbb{R}^{k \times N}$ contains the top k baseline PCs. Values close to 1 indicate that stressed activity largely remained in the same subspace as baseline activity, whereas lower values indicate that activity moved into dimensions that were not dominant at baseline. By construction, $\rho(0) \geq 0.90$.

Using this same baseline PCA subspace, baseline and stressed activity were projected onto the selected baseline PCs. For each trial and time point, we computed the Euclidean distance between the stressed and baseline projected activity vectors from the same trial. These distances were then averaged across trials and time points to obtain one displacement value for each seed, network, and stress condition.

At each stress level, naive and resilient networks were compared using two-sided Mann--Whitney $U$
tests. Comparisons between stress levels within each network class used
two-sided Wilcoxon signed-rank tests, pairing observations from the same
independently trained network by seed. Variance comparisons used
Brown--Forsythe tests; full test statistics are reported in the Supplemental Information,
Table~\hyperref[tab:geometry_stats]{S3}.

\subsection{Metabolic energy analysis}\label{sec:methods_energy_analysis}

We quantified the total per-timestep metabolic energy associated with recurrent
activity and synaptic transmission following~\cite{Ali2022-vw}. For a network
with $N$ recurrent units, we defined based on the network activity vector $r$
\begin{equation}
r =
\begin{pmatrix}
    r_E \\
    r_I
\end{pmatrix},\qquad
\mathcal{R}(t)=\frac{1}{N}\sum_i r_i(t),\qquad
\mathcal{S}(t)=\frac{1}{N}\sum_i\left[
\sum_j |\tilde{W}^{\mathrm{rec}}_{ij}(\delta)|r_j(t)
+\sum_k |W^{\mathrm{in}}_{ik}|u_k(t)\right],
\end{equation}
where $\tilde{W}^{\mathrm{rec}}(\delta)$ is the recurrent weight matrix $W_{\mathrm{rec}}$ after applying
the test-time stress perturbation with magnitude $\delta$. Total energy was
\begin{equation}
E(t)=\frac{1}{3}\mathcal{R}(t)+\frac{2}{3}\mathcal{S}(t).
\end{equation}
We evaluated $E(t)$ on the same 5000 shared trials used for the recurrent
activity analyses and averaged across trials and time to obtain one value per
seed and condition. The full activity and source-to-target synaptic
decomposition is provided in the Supporting Information, S1 Text and S1 Figure.

To compare stress-induced changes while controlling for baseline differences,
we computed the per-seed relative energy deviation between baseline (BL) and stressed with magnitude $\delta$
\begin{equation}
D_E^{\mathrm{seed}}(\delta)
=
\frac{\left|E^{\mathrm{seed}}_{\delta} - E^{\mathrm{seed}}_{\mathrm{BL}}\right|}
{E^{\mathrm{seed}}_{\mathrm{BL}}},
\end{equation}
where $E^{\mathrm{seed}}_{\mathrm{BL}}$ is the unstressed baseline for the same
network. Baseline and stressed total-energy values were compared within each
network class using two-sided Wilcoxon signed-rank tests, pairing observations
from the same network by seed. Independent comparisons between naive and
resilient networks used two-sided Mann--Whitney $U$ tests. Brown--Forsythe tests
compared the between-network variance of relative energy deviation between
classes. Full test statistics are reported in the Supporting Information,
Table~\hyperref[tab:geometry_stats]{S3}.

\subsection{Delay-generalisation trade-off analysis} 

During training, delays between stimuli $S_1$ and $S_2$ were sampled from $400$--$900~\mathrm{ms}$. We therefore defined delays in this interval as within-distribution delays, and delays longer than $900~\mathrm{ms}$ as long out-of-distribution delays, denoted OOD-high. For each trained network, accuracy was averaged separately over within-distribution and OOD-high delays. Accuracy was computed from $2000$ trials per delay condition, excluding zero-coherence trials. A trial was counted as correct when the sign of the time-averaged difference between the network’s two outputs, $y_2(t)-y_1(t)$, during the decision period matched the sign of the stimulus difference, $S_2-S_1$.

We analysed two generalisation trade-off experiments. In the training stress magnitude experiment, resilient networks were trained with maximum stress magnitude
$\delta_{\max} \in \{0.25, 0.5, 0.75, 1.0, 1.5, 2.0\}$.
For each value of $\delta_{\max}$, $50$ independently trained networks were evaluated for each network type, naive and resilient. In the network size experiment, we tested $N \in \{50, 100, 200, 400, 500\}$, with the excitatory population fixed at $N_{\mathrm{E}} = 0.8N$, rounded to the nearest integer. Again, for each network size, $50$ independently trained networks were evaluated for each network type. All networks were evaluated under two test conditions: no stress, and applied test-time stress using $\SItoE$ at $\delta = 0.25$. For these sweeps, we retained the checkpoint with the lowest validation loss and applied early stopping within condition-specific training budgets, resulting in $9000$-$18000$ optimisation steps for the training-stress sweep and $8000$-$40000$ steps across network sizes.

For each experiment, delay region (within- or out-of-distribution), and evaluation condition, we quantified the group difference as
$$
\Delta^a = \bar a_{\mathrm{Naive}} - \bar a_{\mathrm{Resilient}},
$$
where $a$ denotes accuracy. Thus, $\Delta^a > 0$ indicates better performance in Naive networks, whereas $\Delta^a < 0$ indicates better performance in Resilient networks. Because Naive and Resilient networks were trained from independent random initialisations and training seeds, we treated the two network classes as independent replicate groups rather than paired observations. Confidence intervals for $\Delta^a$ were therefore computed using the independent-groups standard error,
\begin{equation}
SE_{\Delta} =
\sqrt{
\frac{s^2_{\mathrm{Naive}}}{n_{\mathrm{Naive}}}
+
\frac{s^2_{\mathrm{Resilient}}}{n_{\mathrm{Resilient}}}
}.
\end{equation}

For each sweep, delay region, and test condition, we treated network type (naive or resilient) as a between-network factor and fitted ordinary least-squares models with heteroscedasticity-consistent HC3 standard errors. The main effect of network type, adjusted for the swept parameter, was tested with $\mathrm{accuracy} \sim \mathrm{network\ type} + \mathrm{parameter}$,
where the swept parameter was $\delta_{\max}$ in the training stress magnitude experiment and $N$ in the network size experiment. To test whether the naive--resilient performance gap changed across each sweep, we fit the interaction model $\mathrm{accuracy} \sim \mathrm{network\ type} \times \mathrm{parameter}$.
The interaction term tested whether the generalisation trade-off varied linearly with stress-training magnitude or network capacity. As direct descriptive comparisons, we also computed independent two-sided Welch tests comparing naive and resilient accuracies while allowing unequal variances between groups, both across the complete sweep and separately at each sweep point.

\subsection{Network density and reciprocity measures}

Graph-theoretical analyses were performed on the effective trained recurrent connectivity matrices, $W_{\mathrm{rec}}$. The matrices were interpreted as weighted, directed, and signed graphs according to the convention $W_{\mathrm{rec}}[\mathrm{post},\mathrm{pre}]$. Density was computed from the binary directed support graph $A_{ij}
=
\mathbf{1}
\left(
|W_{ij}|>0
\right)$, with $ i\neq j,
$ as
$$
D=
\frac{
\sum_{i\neq j}A_{ij}
}{
N(N-1)
}.
$$

Weighted directed reciprocity was computed following \cite{Squartini2013-tw}. Because recurrent E/I connectivity is signed, we adapted the measure to absolute element-wise weight magnitudes, $
W_{ij}^{\mathrm{abs}}
=
|W_{ij}|,
$ with self-connections excluded, as
$$
R_w=
\frac{
\sum_{i\neq j}
\min
\left(
W_{ij}^{\mathrm{abs}},
W_{ji}^{\mathrm{abs}}
\right)
}{
\sum_{i\neq j}
W_{ij}^{\mathrm{abs}}
}.
$$

\subsection{Software and reproducibility}
We implemented all simulations and model training in Python. Differentiable 
recurrent-network simulations and accelerated numerical computations were 
performed using JAX \cite{Bradbury2018-ij}, with parameter optimisation 
implemented through Optax. Numerical analyses, dimensionality reduction,
data handling, and statistical modelling were conducted using NumPy 
\cite{Harris2020-uf}, SciPy \cite{Virtanen2020-ml}, pandas \cite{McKinney2010-yx}, 
scikit-learn \cite{Pedregosa2011-gj}, and statsmodels \cite{Seabold2010-zl}. Figures were 
generated with Matplotlib \cite{Hunter2007-lq}. All analyses were organised as reproducible 
pipeline scripts accompanying the paper. Simulations and model training were run on the GPU workstation of the Institute for Quantitative and Computational Biosciences (IQCB), Johannes Gutenberg University, Mainz, Germany, using NVIDIA L40S GPUs.

\subsection{Use of artificial intelligence tools}

OpenAI ChatGPT and OpenAI Codex were used as assistive tools during the
research and preparation of this article, including idea exploration and
refinement, code development and refactoring, debugging, testing,
documentation, organisation of reproducibility workflows, manuscript
drafting, and language editing. The authors reviewed, revised, and verified
all AI-generated material. AI-assisted code was executed and validated through
tests, smoke workflows, and comparison with the reported analyses and expected
outputs. All scientific decisions, hypotheses, interpretations, results,
claims, conclusions, limitations, implications, and final manuscript
revisions represent the authors' own work and ideas. The authors take full
responsibility for the article, source code, figures, and supporting
materials.

\section*{Author contributions}
Conceptualisation MAD, JH; Data Curation MAD; Formal Analysis MAD, MAB; Funding Acquisition JH; Investigation MAD, MAB; Methodology MAD, JH; Project Administration MAD, JH; Resources MAD; Software MAD, MAB; Supervision JH; Validation MAD; Visualization MAD, MAB; Writing – Original Draft Preparation MAD; Writing – Review \& Editing MAD, MAB, JH.

\section*{Acknowledgements}

The study was funded by the Boeringer Ingelheim Stiftung. We gratefully acknowledge the support of the Institute for Quantitative and Computational Biosciences (IQCB), Johannes Gutenberg University, Mainz, Germany.

\section*{Data and code availability}

The data underlying all figures and statistical analyses are
available from Zenodo at \url{https://doi.org/10.5281/zenodo.21944583}.
The source code, configurations, and instructions required to reproduce the
analyses are available at
\url{https://github.com/mjadiaz/stress-resilience-dale-rnns}.
The version of the code associated with this article is archived at
\url{https://doi.org/10.5281/zenodo.21944893}.

\bibliographystyle{plos2025}
\bibliography{paperpile_bibtex}

@ARTICLE{kalisch_neurobiology_2024,
  title     = "Neurobiology and systems biology of stress resilience",
  author    = "Kalisch, Raffael and Russo, Scott J. and M{\"u}ller, Marianne B.",
  journal   = "Physiological Reviews",
  publisher = "American Physiological Society",
  volume    =  104,
  number    =  3,
  pages     = "1205--1263",
  year      =  2024,
  doi       = "10.1152/physrev.00042.2023"
}

@ARTICLE{Dale1935-yk,
  title     = "Pharmacology and nerve-endings",
  author    = "Dale, Henry",
  journal   = "Proc. R. Soc. Med.",
  publisher = "SAGE Publications",
  volume    =  28,
  number    =  3,
  pages     = "319--332",
  month     =  jan,
  year      =  1935,
  url       = "http://dx.doi.org/10.1177/003591573502800330",
  doi       = "10.1177/003591573502800330",
  language  = "en"
}

@ARTICLE{Squartini2013-tw,
  title     = "Reciprocity of weighted networks",
  author    = "Squartini, Tiziano and Picciolo, Francesco and Ruzzenenti,
               Franco and Garlaschelli, Diego",
  journal   = "Sci. Rep.",
  publisher = "Springer Science and Business Media LLC",
  volume    =  3,
  number    =  1,
  pages     =  2729,
  month     =  sep,
  year      =  2013,
  url       = "http://dx.doi.org/10.1038/srep02729",
  doi       = "10.1038/srep02729",
  language  = "en"
}

@ARTICLE{Hunter2007-lq,
  title     = "Matplotlib: A {2D} graphics environment",
  author    = "Hunter, J D",
  journal   = "Computing in Science \& Engineering",
  publisher = "IEEE COMPUTER SOC",
  volume    =  9,
  number    =  3,
  pages     = "90--95",
  year      =  2007,
  url       = "http://dx.doi.org/10.1109/MCSE.2007.55",
  doi       = "10.1109/MCSE.2007.55"
}

@INPROCEEDINGS{Seabold2010-zl,
  title     = "statsmodels: Econometric and statistical modeling with python",
  author    = "Seabold, Skipper and Perktold, Josef",
  booktitle = "9th Python in Science Conference",
  pages     = "92--96",
  year      =  2010,
  url       = "https://doi.org/10.25080/Majora-92bf1922-011",
  doi       = "10.25080/Majora-92bf1922-011"
}

@ARTICLE{Pedregosa2011-gj,
  title    = "Scikit-learn: Machine Learning in Python",
  author   = "Pedregosa, F and Varoquaux, G and Gramfort, A and Michel, V and
              Thirion, B and Grisel, O and Blondel, M and Prettenhofer, P and
              Weiss, R and Dubourg, V and Vanderplas, J and Passos, A and
              Cournapeau, D and Brucher, M and Perrot, M and Duchesnay, E",
  journal  = "Journal of Machine Learning Research",
  volume   =  12,
  pages    = "2825--2830",
  year     =  2011
}

@INPROCEEDINGS{McKinney2010-yx,
  title     = "Data Structures for Statistical Computing in Python",
  author    = "McKinney, Wes",
  editor    = "van der Walt, Stéfan and Millman, Jarrod",
  booktitle = "Proceedings of the 9th Python in Science Conference",
  pages     = "56--61",
  year      =  2010,
  url       = "http://dx.doi.org/10.25080/Majora-92bf1922-00a",
  doi       = "10.25080/Majora-92bf1922-00a"
}

@ARTICLE{Virtanen2020-ml,
  title    = "{SciPy} 1.0: Fundamental Algorithms for Scientific Computing in
              Python",
  author   = "Virtanen, Pauli and Gommers, Ralf and Oliphant, Travis E and
              Haberland, Matt and Reddy, Tyler and Cournapeau, David and
              Burovski, Evgeni and Peterson, Pearu and Weckesser, Warren and
              Bright, Jonathan and van der Walt, Stéfan J and Brett, Matthew and
              Wilson, Joshua and Millman, K Jarrod and Mayorov, Nikolay and
              Nelson, Andrew R J and Jones, Eric and Kern, Robert and Larson,
              Eric and Carey, C J and Polat, İlhan and Feng, Yu and Moore, Eric
              W and VanderPlas, Jake and Laxalde, Denis and Perktold, Josef and
              Cimrman, Robert and Henriksen, Ian and Quintero, E A and Harris,
              Charles R and Archibald, Anne M and Ribeiro, Antônio H and
              Pedregosa, Fabian and van Mulbregt, Paul and {SciPy 1.0
              Contributors}",
  journal  = "Nature Methods",
  volume   =  17,
  pages    = "261--272",
  year     =  2020,
  url      = "http://dx.doi.org/10.1038/s41592-019-0686-2",
  doi      = "10.1038/s41592-019-0686-2"
}

@ARTICLE{Harris2020-uf,
  title     = "Array programming with {NumPy}",
  author    = "Harris, Charles R and Millman, K Jarrod and van der Walt, Stéfan
               J and Gommers, Ralf and Virtanen, Pauli and Cournapeau, David and
               Wieser, Eric and Taylor, Julian and Berg, Sebastian and Smith,
               Nathaniel J and Kern, Robert and Picus, Matti and Hoyer, Stephan
               and van Kerkwijk, Marten H and Brett, Matthew and Haldane, Allan
               and del Río, Jaime Fernández and Wiebe, Mark and Peterson, Pearu
               and Gérard-Marchant, Pierre and Sheppard, Kevin and Reddy, Tyler
               and Weckesser, Warren and Abbasi, Hameer and Gohlke, Christoph
               and Oliphant, Travis E",
  journal   = "Nature",
  publisher = "Springer Science and Business Media LLC",
  volume    =  585,
  number    =  7825,
  pages     = "357--362",
  month     =  sep,
  year      =  2020,
  url       = "https://doi.org/10.1038/s41586-020-2649-2",
  doi       = "10.1038/s41586-020-2649-2"
}

@MISC{Bradbury2018-ij,
  title    = "{JAX}: composable transformations of {Python+NumPy} programs",
  author   = "Bradbury, James and Frostig, Roy and Hawkins, Peter and Johnson,
              Matthew James and Katariya, Yash and Leary, Chris and Maclaurin,
              Dougal and Necula, George and Paszke, Adam and VanderPlas, Jake
              and Wanderman-Milne, Skye and Zhang, Qiao",
  year     =  2018,
  url      = "http://github.com/jax-ml/jax",
  version  = "0.3.13"
}

@ARTICLE{Compte2000-gw,
  title     = "Synaptic mechanisms and network dynamics underlying spatial
               working memory in a cortical network model",
  author    = "Compte, A and Brunel, N and Goldman-Rakic, P S and Wang, X J",
  journal   = "Cereb. Cortex",
  publisher = "Oxford University Press (OUP)",
  volume    =  10,
  number    =  9,
  pages     = "910--923",
  month     =  sep,
  year      =  2000,
  url       = "http://dx.doi.org/10.1093/cercor/10.9.910",
  doi       = "10.1093/cercor/10.9.910",
  language  = "en"
}

@ARTICLE{Wang1999-kl,
  title     = "Synaptic basis of cortical persistent activity: the importance of
               {NMDA} receptors to working memory",
  author    = "Wang, X J",
  journal   = "J. Neurosci.",
  publisher = "Society for Neuroscience",
  volume    =  19,
  number    =  21,
  pages     = "9587--9603",
  month     =  "1~" # nov,
  year      =  1999,
  url       = "http://dx.doi.org/10.1523/JNEUROSCI.19-21-09587.1999",
  doi       = "10.1523/jneurosci.19-21-09587.1999",
  language  = "en"
}

@ARTICLE{Barak2010-fk,
  title     = "Neuronal population coding of parametric working memory",
  author    = "Barak, Omri and Tsodyks, Misha and Romo, Ranulfo",
  journal   = "J. Neurosci.",
  publisher = "Society for Neuroscience",
  volume    =  30,
  number    =  28,
  pages     = "9424--9430",
  month     =  "14~" # jul,
  year      =  2010,
  url       = "http://dx.doi.org/10.1523/JNEUROSCI.1875-10.2010",
  doi       = "10.1523/JNEUROSCI.1875-10.2010",
  language  = "en"
}

@ARTICLE{Romo1999-la,
  title     = "Neuronal correlates of parametric working memory in the
               prefrontal cortex",
  author    = "Romo, R and Brody, C D and Hernández, A and Lemus, L",
  journal   = "Nature",
  publisher = "Springer Science and Business Media LLC",
  volume    =  399,
  number    =  6735,
  pages     = "470--473",
  month     =  "3~" # jun,
  year      =  1999,
  url       = "http://dx.doi.org/10.1038/20939",
  doi       = "10.1038/20939",
  language  = "en"
}

@ARTICLE{Piwek2023-sg,
  title     = "A recurrent neural network model of prefrontal brain activity
               during a working memory task",
  author    = "Piwek, Emilia P and Stokes, Mark G and Summerfield, Christopher",
  journal   = "PLoS Comput. Biol.",
  publisher = "Public Library of Science (PLoS)",
  volume    =  19,
  number    =  10,
  pages     = "e1011555",
  month     =  "18~" # oct,
  year      =  2023,
  url       = "http://dx.doi.org/10.1371/journal.pcbi.1011555",
  doi       = "10.1371/journal.pcbi.1011555",
  language  = "en"
}

@ARTICLE{Cueva2021-hs,
  title         = "Recurrent neural network models for working memory of
                   continuous variables: activity manifolds, connectivity
                   patterns, and dynamic codes",
  author        = "Cueva, Christopher J and Ardalan, Adel and Tsodyks, Misha and
                   Qian, Ning",
  journal       = "arXiv [q-bio.NC]",
  month         =  "1~" # nov,
  year          =  2021,
  url           = "http://arxiv.org/abs/2111.01275",
  archivePrefix = "arXiv",
  primaryClass  = "q-bio.NC",
  doi           = "10.48550/arXiv.2111.01275"
}

@ARTICLE{Joels2009-gm,
  title     = "The neuro-symphony of stress",
  author    = "Joëls, Marian and Baram, Tallie Z",
  journal   = "Nat. Rev. Neurosci.",
  publisher = "Springer Science and Business Media LLC",
  volume    =  10,
  number    =  6,
  pages     = "459--466",
  month     =  jun,
  year      =  2009,
  url       = "http://dx.doi.org/10.1038/nrn2632",
  doi       = "10.1038/nrn2632",
  language  = "en"
}

@ARTICLE{Du2025-ri,
  title     = "Synapses mediate the effects of different types of stress on
               working memory: a brain-inspired spiking neural network study",
  author    = "Du, Chengcheng and Sun, Yinqian and Wang, Jihang and Zhang, Qian
               and Zeng, Yi",
  journal   = "Front. Cell. Neurosci.",
  publisher = "Frontiers",
  volume    =  19,
  pages     =  1534839,
  month     =  "19~" # mar,
  year      =  2025,
  url       = "http://dx.doi.org/10.3389/fncel.2025.1534839",
  doi       = "10.3389/fncel.2025.1534839",
  language  = "en"
}

@ARTICLE{Roach2023-ti,
  title     = "Choice selective inhibition drives stability and competition in
               decision circuits",
  author    = "Roach, James P and Churchland, Anne K and Engel, Tatiana A",
  journal   = "Nat. Commun.",
  publisher = "Springer Science and Business Media LLC",
  volume    =  14,
  number    =  1,
  pages     =  147,
  month     =  "10~" # jan,
  year      =  2023,
  url       = "http://dx.doi.org/10.1038/s41467-023-35822-8",
  doi       = "10.1038/s41467-023-35822-8",
  language  = "en"
}

@ARTICLE{Lam2022-ly,
  title     = "Effects of altered excitation-inhibition balance on decision
               making in a cortical circuit model",
  author    = "Lam, Norman H and Borduqui, Thiago and Hallak, Jaime and Roque,
               Antonio and Anticevic, Alan and Krystal, John H and Wang,
               Xiao-Jing and Murray, John D",
  journal   = "J. Neurosci.",
  publisher = "Society for Neuroscience",
  volume    =  42,
  number    =  6,
  pages     = "1035--1053",
  month     =  "9~" # feb,
  year      =  2022,
  url       = "http://dx.doi.org/10.1523/JNEUROSCI.1371-20.2021",
  doi       = "10.1523/JNEUROSCI.1371-20.2021",
  language  = "en"
}

@ARTICLE{Nestler2024-jj,
  title     = "Neurobiological basis of stress resilience",
  author    = "Nestler, Eric J and Russo, Scott J",
  journal   = "Neuron",
  publisher = "Elsevier BV",
  volume    =  112,
  number    =  12,
  pages     = "1911--1929",
  month     =  "19~" # jun,
  year      =  2024,
  url       = "http://dx.doi.org/10.1016/j.neuron.2024.05.001",
  doi       = "10.1016/j.neuron.2024.05.001",
  language  = "en"
}

@ARTICLE{Dias-Ferreira2009-nh,
  title     = "Chronic stress causes frontostriatal reorganization and affects
               decision-making",
  author    = "Dias-Ferreira, Eduardo and Sousa, João C and Melo, Irene and
               Morgado, Pedro and Mesquita, Ana R and Cerqueira, João J and
               Costa, Rui M and Sousa, Nuno",
  journal   = "Science",
  publisher = "American Association for the Advancement of Science (AAAS)",
  volume    =  325,
  number    =  5940,
  pages     = "621--625",
  month     =  "31~" # jul,
  year      =  2009,
  url       = "http://dx.doi.org/10.1126/science.1171203",
  doi       = "10.1126/science.1171203",
  language  = "en"
}

@ARTICLE{Arnsten2009-pn,
  title     = "Stress signalling pathways that impair prefrontal cortex
               structure and function",
  author    = "Arnsten, Amy F T",
  journal   = "Nat. Rev. Neurosci.",
  publisher = "Springer Science and Business Media LLC",
  volume    =  10,
  number    =  6,
  pages     = "410--422",
  month     =  jun,
  year      =  2009,
  url       = "http://dx.doi.org/10.1038/nrn2648",
  doi       = "10.1038/nrn2648",
  language  = "en"
}

@ARTICLE{Masse2019-xj,
  title     = "Circuit mechanisms for the maintenance and manipulation of
               information in working memory",
  author    = "Masse, Nicolas Y and Yang, Guangyu R and Song, H Francis and
               Wang, Xiao-Jing and Freedman, David J",
  journal   = "Nat. Neurosci.",
  publisher = "Springer Science and Business Media LLC",
  volume    =  22,
  number    =  7,
  pages     = "1159--1167",
  month     =  "10~" # jul,
  year      =  2019,
  url       = "http://dx.doi.org/10.1038/s41593-019-0414-3",
  doi       = "10.1038/s41593-019-0414-3",
  language  = "en"
}

@ARTICLE{Harris2015-gh,
  title     = "The neocortical circuit: themes and variations",
  author    = "Harris, Kenneth D and Shepherd, Gordon M G",
  journal   = "Nat. Neurosci.",
  publisher = "Springer Science and Business Media LLC",
  volume    =  18,
  number    =  2,
  pages     = "170--181",
  month     =  feb,
  year      =  2015,
  url       = "http://dx.doi.org/10.1038/nn.3917",
  doi       = "10.1038/nn.3917",
  language  = "en"
}

@ARTICLE{Tsuda2026-tl,
  title     = "Neuromodulators generate multiple context-relevant behaviors in
               recurrent neural networks",
  author    = "Tsuda, Ben and Pate, Stefan C and Tye, Kay M and Siegelmann, Hava
               T and Sejnowski, Terrence J",
  journal   = "Neural Comput.",
  publisher = "MIT Press",
  volume    =  38,
  number    =  3,
  pages     = "292--327",
  month     =  "27~" # feb,
  year      =  2026,
  url       = "https://dx.doi.org/10.1162/NECO.a.1489",
  doi       = "10.1162/NECO.a.1489",
  language  = "en"
}

@ARTICLE{Ali2022-vw,
  title     = "Predictive coding is a consequence of energy efficiency in
               recurrent neural networks",
  author    = "Ali, Abdullahi and Ahmad, Nasir and de Groot, Elgar and Johannes
               van Gerven, Marcel Antonius and Kietzmann, Tim Christian",
  journal   = "Patterns (N. Y.)",
  publisher = "Elsevier BV",
  volume    =  3,
  number    =  12,
  pages     =  100639,
  month     =  "9~" # dec,
  year      =  2022,
  url       = "http://dx.doi.org/10.1016/j.patter.2022.100639",
  doi       = "10.1016/j.patter.2022.100639",
  language  = "en"
}

@ARTICLE{Wu2020-mt,
  title         = "Adversarial Weight Perturbation helps robust generalization",
  author        = "Wu, Dongxian and Xia, Shu-Tao and Wang, Yisen",
  journal       = "arXiv [cs.LG]",
  month         =  "13~" # apr,
  year          =  2020,
  url           = "http://dx.doi.org/10.5555/3495724.3495973",
  archivePrefix = "arXiv",
  primaryClass  = "cs.LG",
  doi           = "10.5555/3495724.3495973",
  language      = "en"
}

@ARTICLE{Rungratsameetaweemana2025-cc,
  title     = "Random noise promotes slow heterogeneous synaptic dynamics
               important for robust working memory computation",
  author    = "Rungratsameetaweemana, Nuttida and Kim, Robert and Chotibut,
               Thiparat and Sejnowski, Terrence J",
  journal   = "Proc. Natl. Acad. Sci. U. S. A.",
  publisher = "Proceedings of the National Academy of Sciences",
  volume    =  122,
  number    =  3,
  pages     = "e2316745122",
  month     =  "21~" # jan,
  year      =  2025,
  url       = "http://dx.doi.org/10.1073/pnas.2316745122",
  doi       = "10.1073/pnas.2316745122",
  language  = "en"
}

@ARTICLE{Lim2021-yy,
  title    = "Noisy Recurrent Neural Networks",
  author   = "Lim, Soon Hoe and Erichson, N Benjamin and Hodgkinson, Liam and
              Mahoney, Michael W",
  journal  = "Advances in Neural Information Processing Systems",
  volume   =  34,
  pages    = "5124--5137",
  month    =  "6~" # dec,
  year     =  2021,
  url      = "https://proceedings.neurips.cc/paper/2021/hash/29301521774ff3cbd26652b2d5c95996-Abstract.html"
}

@ARTICLE{Sears2021-lz,
  title     = "Influence of glutamate and {GABA} transport on brain
               excitatory/inhibitory balance",
  author    = "Sears, Sheila Ms and Hewett, Sandra J",
  journal   = "Exp. Biol. Med. (Maywood)",
  publisher = "SAGE Publications",
  volume    =  246,
  number    =  9,
  pages     = "1069--1083",
  month     =  "7~" # may,
  year      =  2021,
  url       = "http://dx.doi.org/10.1177/1535370221989263",
  doi       = "10.1177/1535370221989263",
  language  = "en"
}

@ARTICLE{Girotti2024-vt,
  title     = "Effects of chronic stress on cognitive function - From
               neurobiology to intervention",
  author    = "Girotti, Milena and Bulin, Sarah E and Carreno, Flavia R",
  journal   = "Neurobiol. Stress",
  publisher = "Elsevier BV",
  volume    =  33,
  number    =  100670,
  pages     =  100670,
  month     =  "1~" # nov,
  year      =  2024,
  url       = "http://dx.doi.org/10.1016/j.ynstr.2024.100670",
  doi       = "10.1016/j.ynstr.2024.100670",
  language  = "en"
}

@ARTICLE{McEwen2016-wi,
  title     = "Stress effects on neuronal structure: Hippocampus, amygdala, and
               prefrontal cortex",
  author    = "McEwen, Bruce S and Nasca, Carla and Gray, Jason D",
  journal   = "Neuropsychopharmacology",
  publisher = "Springer Science and Business Media LLC",
  volume    =  41,
  number    =  1,
  pages     = "3--23",
  month     =  jan,
  year      =  2016,
  url       = "http://dx.doi.org/10.1038/npp.2015.171",
  doi       = "10.1038/npp.2015.171",
  language  = "en"
}

@ARTICLE{Kim2023-yu,
  title     = "Neurocognitive effects of stress: a metaparadigm perspective",
  author    = "Kim, Eun Joo and Kim, Jeansok J",
  journal   = "Mol. Psychiatry",
  publisher = "Nature Publishing Group",
  volume    =  28,
  number    =  7,
  pages     = "2750--2763",
  month     =  "9~" # jul,
  year      =  2023,
  url       = "http://dx.doi.org/10.1038/s41380-023-01986-4",
  doi       = "10.1038/s41380-023-01986-4",
  language  = "en"
}

@ARTICLE{Godoy2018-js,
  title     = "A comprehensive overview on stress neurobiology: Basic concepts
               and clinical implications",
  author    = "Godoy, Lívea Dornela and Rossignoli, Matheus Teixeira and
               Delfino-Pereira, Polianna and Garcia-Cairasco, Norberto and de
               Lima Umeoka, Eduardo Henrique",
  journal   = "Front. Behav. Neurosci.",
  publisher = "Frontiers",
  volume    =  12,
  pages     =  127,
  month     =  "3~" # jul,
  year      =  2018,
  url       = "http://dx.doi.org/10.3389/fnbeh.2018.00127",
  doi       = "10.3389/fnbeh.2018.00127",
  language  = "en"
}

@ARTICLE{Kirischuk2022-ml,
  title     = "Keeping excitation-inhibition ratio in balance",
  author    = "Kirischuk, Sergei",
  journal   = "Int. J. Mol. Sci.",
  publisher = "MDPI AG",
  volume    =  23,
  number    =  10,
  pages     =  5746,
  month     =  "20~" # may,
  year      =  2022,
  url       = "http://dx.doi.org/10.3390/ijms23105746",
  doi       = "10.3390/ijms23105746",
  language  = "en"
}

@ARTICLE{Driscoll2024-rd,
  title     = "Flexible multitask computation in recurrent networks utilizes
               shared dynamical motifs",
  author    = "Driscoll, Laura N and Shenoy, Krishna and Sussillo, David",
  journal   = "Nat. Neurosci.",
  publisher = "Springer Science and Business Media LLC",
  volume    =  27,
  number    =  7,
  pages     = "1349--1363",
  month     =  "9~" # jul,
  year      =  2024,
  url       = "http://dx.doi.org/10.1038/s41593-024-01668-6",
  doi       = "10.1038/s41593-024-01668-6",
  language  = "en"
}

@ARTICLE{Svensson2018-xd,
  title     = "General principles of neuronal co-transmission: Insights from
               multiple model systems",
  author    = "Svensson, Erik and Apergis-Schoute, John and Burnstock, Geoffrey
               and Nusbaum, Michael P and Parker, David and Schiöth, Helgi B",
  journal   = "Front. Neural Circuits",
  publisher = "Frontiers Media SA",
  volume    =  12,
  pages     =  117,
  year      =  2018,
  url       = "http://dx.doi.org/10.3389/fncir.2018.00117",
  doi       = "10.3389/fncir.2018.00117",
  language  = "en"
}

@ARTICLE{Strata1999-fv,
  title     = "Dale's principle",
  author    = "Strata, P and Harvey, R",
  journal   = "Brain Res. Bull.",
  publisher = "Brain Res Bull",
  volume    =  50,
  number    = "5-6",
  pages     = "349--350",
  month     =  nov,
  year      =  1999,
  url       = "http://dx.doi.org/10.1016/s0361-9230(99)00100-8",
  doi       = "10.1016/s0361-9230(99)00100-8",
  language  = "en"
}

@ARTICLE{Barranca2022-aj,
  title     = "Functional implications of Dale's law in balanced neuronal
               network dynamics and decision making",
  author    = "Barranca, Victor J and Bhuiyan, Asha and Sundgren, Max and Xing,
               Fangzhou",
  journal   = "Front. Neurosci.",
  publisher = "Frontiers Media SA",
  volume    =  16,
  pages     =  801847,
  month     =  "28~" # feb,
  year      =  2022,
  url       = "http://dx.doi.org/10.3389/fnins.2022.801847",
  doi       = "10.3389/fnins.2022.801847",
  language  = "en"
}

@ARTICLE{Ingrosso2019-mr,
  title     = "Training dynamically balanced excitatory-inhibitory networks",
  author    = "Ingrosso, Alessandro and Abbott, L F",
  journal   = "PLoS One",
  publisher = "Public Library of Science (PLoS)",
  volume    =  14,
  number    =  8,
  pages     = "e0220547",
  month     =  "8~" # aug,
  year      =  2019,
  url       = "http://dx.doi.org/10.1371/journal.pone.0220547",
  doi       = "10.1371/journal.pone.0220547",
  language  = "en"
}

@ARTICLE{Jarne2024-mm,
  title     = "Effect in the spectra of eigenvalues and dynamics of {RNNs}
               trained with excitatory-inhibitory constraint",
  author    = "Jarne, Cecilia and Caruso, Mariano",
  journal   = "Cogn. Neurodyn.",
  publisher = "Springer Science and Business Media LLC",
  volume    =  18,
  number    =  3,
  pages     = "1323--1335",
  month     =  jun,
  year      =  2024,
  url       = "http://dx.doi.org/10.1007/s11571-023-09956-w",
  doi       = "10.1007/s11571-023-09956-w",
  language  = "en"
}

@ARTICLE{Ma2020-to,
  title         = "A neural network walks into a lab: towards using deep nets as
                   models for human behavior",
  author        = "Ma, Wei Ji and Peters, Benjamin",
  journal       = "arXiv [cs.AI]",
  month         =  "2~" # may,
  year          =  2020,
  url           = "http://arxiv.org/abs/2005.02181",
  archivePrefix = "arXiv",
  primaryClass  = "cs.AI",
  doi           = "10.48550/arXiv.2005.02181"
}

@ARTICLE{Sussillo2013-fg,
  title     = "Opening the black box: low-dimensional dynamics in
               high-dimensional recurrent neural networks",
  author    = "Sussillo, David and Barak, Omri",
  journal   = "Neural Comput.",
  publisher = "MIT Press - Journals",
  volume    =  25,
  number    =  3,
  pages     = "626--649",
  month     =  mar,
  year      =  2013,
  url       = "http://dx.doi.org/10.1162/NECO_a_00409",
  doi       = "10.1162/NECO\_a\_00409",
  language  = "en"
}

@ARTICLE{Kao2019-ns,
  title     = "Considerations in using recurrent neural networks to probe neural
               dynamics",
  author    = "Kao, Jonathan C",
  journal   = "J. Neurophysiol.",
  publisher = "American Physiological Society",
  volume    =  122,
  number    =  6,
  pages     = "2504--2521",
  month     =  "1~" # dec,
  year      =  2019,
  url       = "http://dx.doi.org/10.1152/jn.00467.2018",
  doi       = "10.1152/jn.00467.2018",
  language  = "en"
}

@MISC{Dubreuil2021-sf,
  title        = "Dynamical system approach to explainability in recurrent
                  neural networks",
  author       = "Dubreuil, Alexis",
  year         =  2021,
  howpublished = "\url{https://www.semanticscholar.org/paper/Dynamical-system-approach-to-explainability-in-Dubreuil/d08515aefa1a9df330a33106e788e21daa14c135}",
  note         = "Accessed: 2026-1-16",
  language     = "en"
}

@ARTICLE{Rajakumar2021-uh,
  title     = "Stimulus-driven and spontaneous dynamics in excitatory-inhibitory
               recurrent neural networks for sequence representation",
  author    = "Rajakumar, Alfred and Rinzel, John and Chen, Zhe S",
  journal   = "Neural Comput.",
  publisher = "MIT Press - Journals",
  volume    =  33,
  number    =  10,
  pages     = "2603--2645",
  month     =  "16~" # sep,
  year      =  2021,
  url       = "http://dx.doi.org/10.1162/neco_a_01418",
  doi       = "10.1162/neco\_a\_01418",
  language  = "en"
}

@ARTICLE{Page2019-xg,
  title     = "Prefrontal excitatory/inhibitory balance in stress and emotional
               disorders: Evidence for over-inhibition",
  author    = "Page, Chloe E and Coutellier, Laurence",
  journal   = "Neurosci. Biobehav. Rev.",
  publisher = "Elsevier BV",
  volume    =  105,
  pages     = "39--51",
  month     =  "1~" # oct,
  year      =  2019,
  url       = "http://dx.doi.org/10.1016/j.neubiorev.2019.07.024",
  doi       = "10.1016/j.neubiorev.2019.07.024",
  language  = "en"
}

@ARTICLE{McKlveen2016-px,
  title     = "Chronic stress increases prefrontal inhibition: A mechanism for
               stress-induced prefrontal dysfunction",
  author    = "McKlveen, Jessica M and Morano, Rachel L and Fitzgerald, Maureen
               and Zoubovsky, Sandra and Cassella, Sarah N and Scheimann, Jessie
               R and Ghosal, Sriparna and Mahbod, Parinaz and Packard, Benjamin
               A and Myers, Brent and Baccei, Mark L and Herman, James P",
  journal   = "Biol. Psychiatry",
  publisher = "Elsevier",
  volume    =  80,
  number    =  10,
  pages     = "754--764",
  month     =  "15~" # nov,
  year      =  2016,
  url       = "http://dx.doi.org/10.1016/j.biopsych.2016.03.2101",
  doi       = "10.1016/j.biopsych.2016.03.2101",
  language  = "en"
}

@ARTICLE{Rodrigues2024-ry,
  title     = "Chronic stress alters synaptic inhibition/excitation balance of
               pyramidal neurons but not {PV} interneurons in the infralimbic
               and prelimbic cortices of {C57BL}/{6J} mice",
  author    = "Rodrigues, Diana and Santa, Cátia and Manadas, Bruno and
               Monteiro, Patrícia",
  journal   = "eNeuro",
  publisher = "Society for Neuroscience",
  volume    =  11,
  number    =  8,
  pages     = "ENEURO.0053--24.2024",
  month     =  "1~" # aug,
  year      =  2024,
  url       = "http://dx.doi.org/10.1523/ENEURO.0053-24.2024",
  doi       = "10.1523/ENEURO.0053-24.2024",
  language  = "en"
}

@ARTICLE{Yang2020-hj,
  title     = "Artificial neural networks for neuroscientists: A Primer",
  author    = "Yang, Guangyu Robert and Wang, Xiao-Jing",
  journal   = "Neuron",
  publisher = "Elsevier BV",
  volume    =  107,
  number    =  6,
  pages     = "1048--1070",
  month     =  "23~" # sep,
  year      =  2020,
  url       = "http://dx.doi.org/10.1016/j.neuron.2020.09.005",
  doi       = "10.1016/j.neuron.2020.09.005",
  language  = "en"
}

@ARTICLE{Yang2019-sq,
  title     = "Task representations in neural networks trained to perform many
               cognitive tasks",
  author    = "Yang, Guangyu Robert and Joglekar, Madhura R and Song, H Francis
               and Newsome, William T and Wang, Xiao-Jing",
  journal   = "Nat. Neurosci.",
  publisher = "Springer Science and Business Media LLC",
  volume    =  22,
  number    =  2,
  pages     = "297--306",
  month     =  "14~" # feb,
  year      =  2019,
  url       = "http://dx.doi.org/10.1038/s41593-018-0310-2",
  doi       = "10.1038/s41593-018-0310-2",
  language  = "en"
}

@ARTICLE{Song2016-qj,
  title     = "Training excitatory-inhibitory recurrent neural networks for
               cognitive tasks: A simple and flexible framework",
  author    = "Song, H Francis and Yang, Guangyu R and Wang, Xiao-Jing",
  editor    = "Sporns, Olaf",
  journal   = "PLoS Comput. Biol.",
  publisher = "Public Library of Science (PLoS)",
  volume    =  12,
  number    =  2,
  pages     = "e1004792",
  month     =  "29~" # feb,
  year      =  2016,
  url       = "http://dx.doi.org/10.1371/journal.pcbi.1004792",
  doi       = "10.1371/journal.pcbi.1004792",
  language  = "en"
}

\newpage
\section*{Supplemental Information}
\raggedbottom
\paragraph*{S1 Text.} 
\textbf{Decomposition of the metabolic-energy budget.}

The main analysis uses only total energy, but we additionally decomposed that
total into eight non-overlapping terms. Activity cost was separated by
population,
\begin{equation}
\mathcal{R}_B(t)=\frac{1}{N}\sum_{i\in B}r_i(t),
\qquad B\in\{E,I\}.
\end{equation}
For recurrent transmission, source population $A$ and target population $B$
defined
\begin{equation}
\mathcal{S}_{A\rightarrow B}(t)=\frac{1}{N}
\sum_{i\in B}\sum_{j\in A}
\left|W^{\mathrm{eff}}_{ij}(\delta)\right|r_j(t),
\qquad A,B\in\{E,I\},
\end{equation}
and the external-input contribution to each target population was
\begin{equation}
\mathcal{S}_{\mathrm{ext}\rightarrow B}(t)=\frac{1}{N}
\sum_{i\in B}\sum_k\left|W^{\mathrm{in}}_{ik}\right|u_k(t).
\end{equation}
The components sum exactly to the total used in the main text:
\begin{align}
E(t)={}&\frac{1}{3}\left(\mathcal{R}_E+\mathcal{R}_I\right)
+\frac{2}{3}\bigl(
\mathcal{S}_{E\rightarrow E}+\mathcal{S}_{I\rightarrow E}
+\mathcal{S}_{E\rightarrow I}+\mathcal{S}_{I\rightarrow I}\notag\\
&\hspace{5.5em}+\mathcal{S}_{\mathrm{ext}\rightarrow E}
+\mathcal{S}_{\mathrm{ext}\rightarrow I}\bigr).
\end{align}
Each term was averaged over the same trials and time points as total energy.
For descriptive component-wise comparisons, relative change was computed
within seed as $(Q_\delta-Q_{\mathrm{BL}})/Q_{\mathrm{BL}}$ comparing stress $\delta$ with unstressed baseline (BL); negative values
therefore indicate a reduction under stress. The largest class-specific change
occurred in $\mathcal{S}_{I\rightarrow E}$: its mean contribution decreased by
$29.8\%$ and $40.3\%$ in naive networks at $\delta=0.25$ and $0.5$, respectively,
but increased by $2.3\%$ and $5.5\%$ in resilient networks (Table~\hyperref[tab:geometry_stats]{S3}).

\begin{figure}[H]
\centering
\includegraphics[width=\linewidth]{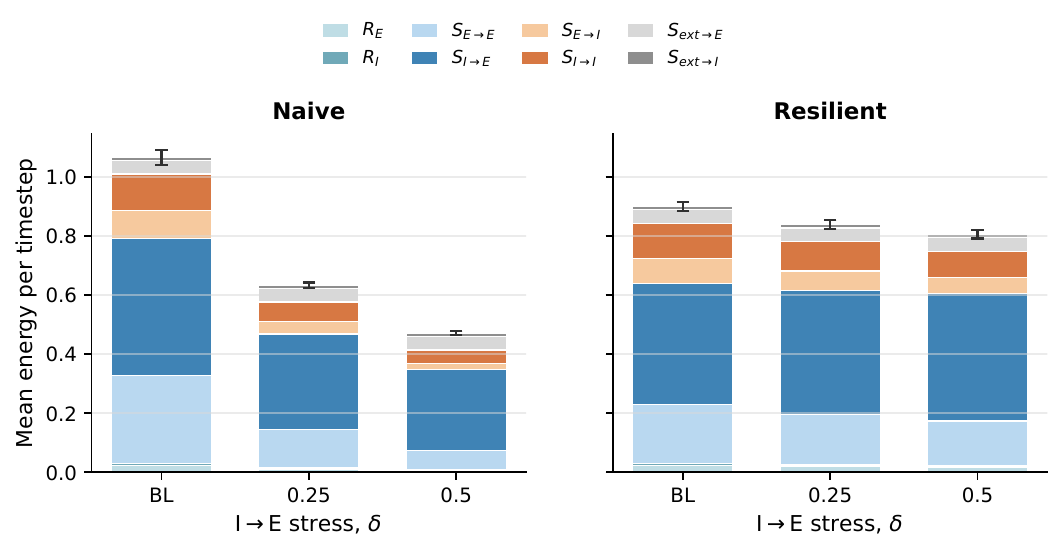}
\caption*{\textbf{S1 Figure. Decomposition of the metabolic-energy budget.}
Stacked bars show the mean contribution of excitatory and inhibitory activity,
the four recurrent source-to-target pathways, and external input to excitatory
and inhibitory targets. Segment heights include the $1/3$ activity or $2/3$
synaptic coefficient, so each stack equals the total energy reported in the
main analysis. Bars compare unstressed baseline (BL) with $I\!\rightarrow\!E$
stress at $\delta=0.25$ and $0.5$ in naive and resilient networks; error bars
show the standard error of the mean (SEM) of total energy across 50 independently trained networks.}
\end{figure}

\paragraph*{S2 Text.}
\textbf{Detailed stress-operator comparison.}

\begingroup
\small
The comparison comprised four targeted perturbations and four matched
population-agnostic controls. Targeted weight perturbations strengthened
inhibitory-to-excitatory weights ($\SItoE$) or weakened recurrent excitatory
weights ($\SEE$); targeted activity perturbations increased inhibitory gain
($\SrI$) or decreased excitatory gain ($\SrE$). The controls applied the same
up- or down-scaling to the complete recurrent matrix ($\SWup$, $\SWdown$) or to
all recurrent activities ($\SrUp$, $\SrDown$).

For weight perturbations, independent factors $m_{ij}$ generated by
Eqs.~\ref{eq:stress_operator}--\ref{eq:stress_params} acted on the selected
submatrix $W_P$,
\begin{equation}
\widetilde W_{P,ij}=m_{ij}W_{P,ij},
\qquad
\mathbb{E}[\widetilde W_{P,ij}]=(1+\delta)W_{P,ij}.
\end{equation}
For activity perturbations, one independently sampled factor $g_{P,j}$ acted on
each neuron in the selected population,
\begin{equation}
r^{\mathrm{eff}}_{P,j}(t)=g_{P,j}r_{P,j}(t),
\qquad
\mathbb{E}[r^{\mathrm{eff}}_{P,j}(t)]=(1+\delta)r_{P,j}(t).
\end{equation}
Thus, a weight operator modified only the selected synapses, whereas an
activity-gain operator changed a neuron's contribution to every postsynaptic
target.

Each operator was evaluated in naive networks and in resilient networks trained
with that same operator. We used 50 independently trained networks per group and
three evaluation magnitudes: unstressed baseline ($\delta=0$), the maximum
training magnitude ($\delta_T=0.5$), and a stronger test beyond the training
range ($\delta_{\max}=1.0$), with the direction reversed for down-scaling
protocols. For every network and magnitude, 100 task trials were each evaluated
under 100 independent stress realisations. Performance measures were accuracy,
readout MSE, and decision confidence, defined as the winner-minus-runner-up
output margin 
averaged over the decision epoch. Dynamics measures were the
time-averaged population activities $\bar r_E$ and $\bar r_I$, recurrent drives
$h_{E\to E}$, $|h_{I\to E}|$, $h_{E\to I}$, and $|h_{I\to I}|$, and the
time-averaged drive ratio $E/|I|=h_{E\to E}/|h_{I\to E}|$.

For this operator-screening analysis, dynamic measures were first averaged over
time and accuracy over stress realisations. We then took the median across
trials for each seed and expressed each stressed summary as a signed percentage
change from that seed's unstressed baseline using
Eq.~\ref{eq:pct_change}.
Fig.~\ref{fig:stress_operator_comparison} reports the median across seeds in
each split cell, with the upper-left triangle showing $\delta_T$ and the
lower-right triangle $\delta_{\max}$. For inferential comparisons, the per-seed
changes were tested against zero using two-sided one-sample Wilcoxon signed-rank
tests, with Benjamini--Hochberg correction across the reported comparisons.
\endgroup

\clearpage

\paragraph*{S1 Table.}
\textbf{Key psychometric accuracy comparisons.}
\phantomsection\label{tab:psychometric_class_stats}

\begin{table}[H]
\centering
\footnotesize
\setlength{\tabcolsep}{3.2pt}
\renewcommand{\arraystretch}{1.15}
\resizebox{\textwidth}{!}{%
\begin{tabular}{@{}lcccccc@{}}
\toprule
\textbf{Comparison} & \textbf{Naive} & \textbf{Resilient} & $\boldsymbol{\Delta}$ & \textbf{95\% CI} & $\boldsymbol{t}$\textbf{ (df)} & \textbf{Adjusted }$\boldsymbol{p}$ \\
\midrule
Curve-wide, unstressed & $0.8861\pm0.0022$ & $0.8878\pm0.0019$ & $-0.0017$ & $[-0.0075,+0.0040]$ & $-0.60$ ($95.9$) & $0.553$ \\
Curve-wide, stress $\delta=0.25$ & $0.7141\pm0.0078$ & $0.8832\pm0.0014$ & $-0.1691$ & $[-0.1850,-0.1533]$ & $-21.39$ ($52.1$) & $<10^{-26}$ \\
Curve-wide stress effect & $-0.1720\pm0.0081$ & $-0.0046\pm0.0009$ & $-0.1674$ & $[-0.1837,-0.1511]$ & $-20.64$ ($50.3$) & $<10^{-25}$ \\
\midrule
$|\Delta|=0.032$, unstressed & $0.8349\pm0.0091$ & $0.8380\pm0.0083$ & $-0.0031$ & $[-0.0276,+0.0214]$ & $-0.25$ ($97.1$) & $1.000$ \\
$|\Delta|=0.032$, stress $\delta=0.25$ & $0.5276\pm0.0037$ & $0.8221\pm0.0060$ & $-0.2945$ & $[-0.3086,-0.2805]$ & $-41.72$ ($80.6$) & $<10^{-50}$ \\
$|\Delta|=0.128$, stress $\delta=0.25$ & $0.6440\pm0.0135$ & $0.9993\pm0.0002$ & $-0.3553$ & $[-0.3823,-0.3282]$ & $-26.38$ ($49.0$) & $<10^{-28}$ \\
\bottomrule
\end{tabular}}
\caption*{Values are mean$\pm$SEM across 50 independently trained networks per class. $\Delta$ is naive minus resilient. Curve-wide values give each tested evidence magnitude equal weight; the stress-effect row is stressed minus unstressed accuracy. Class comparisons use independent two-sided Welch tests. Reported $p$-values are Holm-adjusted within the corresponding comparison family.}
\end{table}

\paragraph*{S2 Table.}
\textbf{Key delay-generalisation comparisons.}
\phantomsection\label{tab:figure1_delay_stats}

\begin{table}[H]
\centering
\footnotesize
\setlength{\tabcolsep}{3.5pt}
\renewcommand{\arraystretch}{1.18}
\resizebox{\textwidth}{!}{%
\begin{tabular}{@{}clcccccc@{}}
\toprule
\textbf{Test }$\boldsymbol{\delta}$ & \textbf{Delay region} & \textbf{Naive} & \textbf{Resilient} & $\boldsymbol{\Delta}$ & \textbf{95\% CI} & \textbf{Welch }$\boldsymbol{t}$\textbf{ (df)} & $\boldsymbol{p}$ \\
\midrule
$0$ & Within ($400$--$900$ ms) & $0.9593\pm0.0030$ & $0.9592\pm0.0026$ & $+0.00004$ & $[-0.00777,+0.00785]$ & $0.01$ ($96.3$) & $0.992$ \\
$0.25$ & Within ($400$--$900$ ms) & $0.6941\pm0.0098$ & $0.9551\pm0.0018$ & $-0.2610$ & $[-0.2810,-0.2410]$ & $-26.24$ ($52.2$) & $<10^{-30}$ \\
$0$ & OOD-high ($1200$--$1800$ ms) & $0.7878\pm0.0229$ & $0.6726\pm0.0259$ & $+0.1152$ & $[+0.0466,+0.1838]$ & $3.33$ ($96.6$) & $0.00122$ \\
$0.25$ & OOD-high ($1200$--$1800$ ms) & $0.6063\pm0.0076$ & $0.6948\pm0.0245$ & $-0.0885$ & $[-0.1398,-0.0372]$ & $-3.45$ ($58.5$) & $0.00104$ \\
\bottomrule
\end{tabular}}
\caption*{For each network, accuracy was averaged over the exact delays in the indicated region; entries are mean$\pm$SEM across 50 independently trained networks per class. $\Delta$ is naive minus resilient. Class comparisons use independent two-sided Welch tests.}
\end{table}

\clearpage

\paragraph*{S3 Table.}
\textbf{Full statistics for the recurrent dynamics and geometry analysis.}
\phantomsection\label{tab:geometry_stats}
\begin{table}[H]
\centering
\footnotesize
\setlength{\tabcolsep}{4pt}
\renewcommand{\arraystretch}{1.2}
\begin{tabular}{@{}p{2.0cm}p{3.0cm}p{2.7cm}llc@{}}
\hline
\textbf{Measure} & \textbf{Comparison} & \textbf{Values} & \textbf{Test} & \textbf{Statistic} & $\boldsymbol{p}$ \\
\hline
Trajectory displacement & Naive vs Resilient, $\delta=0.25$ & $1.761$ vs.\ $0.543$ & MWU & $U=2500$ & $7.07\times10^{-18}$ \\
 & Naive vs Resilient, $\delta=0.5$ & $2.466$ vs.\ $0.890$ & MWU & $U=2500$ & $7.07\times10^{-18}$ \\
 & Naive, $\delta=0.25$ vs.\ $0.5$ & $1.761$ vs.\ $2.466$ & WSR & $W=0$ & $1.78\times10^{-15}$ \\
 & Resilient, $\delta=0.25$ vs.\ $0.5$ & $0.543$ vs.\ $0.890$ & WSR & $W=0$ & $1.78\times10^{-15}$ \\
 & Resilient $\delta=0.5$ vs.\ Naive $\delta=0.25$ & $0.890$ vs.\ $1.761$ & MWU & $U=7$ & $1.08\times10^{-17}$ \\
\hline
Subspace preservation $\rho$ & Naive vs Resilient, $\delta=0.25$ & $0.888$ vs.\ $0.900$ & MWU & $U=918$ & $0.0223$ \\
 & Naive vs Resilient, $\delta=0.5$ & $0.898$ vs.\ $0.883$ & MWU & $U=1542$ & $0.0445$ \\
 & Naive, $\delta=0.25$ vs.\ $0.5$ & $0.888$ vs.\ $0.898$ & WSR & $W=281$ & $3.99\times10^{-4}$ \\
 & Resilient, $\delta=0.25$ vs.\ $0.5$ & $0.900$ vs.\ $0.883$ & WSR & $W=0$ & $1.78\times10^{-15}$ \\
 & Resilient $\delta=0.5$ vs.\ Naive $\delta=0.25$ & $0.883$ vs.\ $0.888$ & MWU & $U=1123$ & $0.383$ \\
\hline
Total energy $E$ & Naive vs Resilient, baseline & $1.066$ vs.\ $0.900$ & MWU & $U=1938$ & $2.14\times10^{-6}$ \\
 & Naive, baseline vs.\ $\delta=0.25$ & $1.066$ vs.\ $0.633$ & WSR & $W=0$ & $1.78\times10^{-15}$ \\
 & Naive, baseline vs.\ $\delta=0.5$ & $1.066$ vs.\ $0.471$ & WSR & $W=0$ & $1.78\times10^{-15}$ \\
 & Resilient, baseline vs.\ $\delta=0.25$ & $0.900$ vs.\ $0.838$ & WSR & $W=0$ & $1.78\times10^{-15}$ \\
 & Resilient, baseline vs.\ $\delta=0.5$ & $0.900$ vs.\ $0.805$ & WSR & $W=0$ & $1.78\times10^{-15}$ \\
\hline
Relative energy deviation & Naive vs Resilient, $\delta=0.25$ & $0.398$ vs.\ $0.068$ & MWU & $U=2500$ & $7.07\times10^{-18}$ \\
 & Naive vs Resilient, $\delta=0.5$ & $0.550$ vs.\ $0.104$ & MWU & $U=2500$ & $7.07\times10^{-18}$ \\
\hline
Energy deviation variance & Naive vs Resilient, $\delta=0.25$ & $\sigma=0.060$ vs.\ $0.019$ & BF & $W=37.08$ & $2.22\times10^{-8}$ \\
 & Naive vs Resilient, $\delta=0.5$ & $\sigma=0.064$ vs.\ $0.031$ & BF & $W=22.87$ & $6.10\times10^{-6}$ \\
\hline
Inhibitory synaptic cost $S_{I\rightarrow E}$ & Naive vs Resilient (change vs.\ baseline, $\delta=0.25/0.5$) & $-29.8\%/{-}40.3\%$ vs.\ $+2.3\%/{+}5.5\%$ & MWU & $U=0/0$ & $7.07\times10^{-18}/7.07\times10^{-18}$ \\
\hline
\end{tabular}
\caption*{
Values give the compared quantities (naive vs.\ resilient for cross-group
rows; $\delta=0.25$ vs.\ $\delta=0.5$ or baseline vs.\ stress for
within-network rows). MWU, Mann--Whitney $U$ test; WSR, paired Wilcoxon
signed-rank test; BF, Brown--Forsythe test (variance). For energy-deviation
variance, values are standard deviations across seeds. The four matched-stress
displacement and subspace comparisons shown in Fig.~\ref{fig:figure2} use
Holm-adjusted $p$ values; the corresponding adjusted values are
$2.83\times10^{-17}$, $2.83\times10^{-17}$, $0.0446$, and $0.0446$.}
\end{table}

\clearpage
\paragraph*{S4 Table.}
\textbf{Full statistics for the delay-generalisation trade-off analysis.}
\phantomsection\label{tab:delay_tradeoff_stats}

\begin{table}[H]
\centering
\footnotesize
\setlength{\tabcolsep}{2.5pt}
\renewcommand{\arraystretch}{1.25}
\begin{tabular}{@{}lllccccl@{}}
\hline
\textbf{Sweep} & \textbf{Delay regime} & \textbf{Evaluation} & \textbf{Naive} & \textbf{Resilient} & $\boldsymbol{\Delta}^a$ & \textbf{Welch} $\boldsymbol{p}$ & \textbf{Interaction} ($\boldsymbol{\beta}$, $\boldsymbol{p}$) \\
\hline
$\delta_T$ & Within-dist. & No stress            & $0.9865$ & $0.9862$ & $+0.0003$ & $0.353$ & $\beta=-0.000801$, $p=0.134$ \\
$\delta_T$ & Within-dist. & Stress $\delta=0.25$ & $0.6886$ & $0.9821$ & $-0.2935$ & $2.81\times10^{-213}$ & $\beta=-0.00750$, $p=0.208$ \\
$\delta_T$ & OOD-high     & No stress            & $0.9177$ & $0.7517$ & $+0.1660$ & $6.44\times10^{-41}$ & $\beta=+0.00273$, $p=0.881$ \\
$\delta_T$ & OOD-high     & Stress $\delta=0.25$ & $0.6300$ & $0.7785$ & $-0.1486$ & $7.80\times10^{-44}$ & $\beta=-0.0194$, $p=0.206$ \\
\hline
$N$ & Within-dist. & No stress            & $0.9851$ & $0.9848$ & $+0.0003$ & $0.639$ & $\beta=+1.23\times10^{-5}$, $p=3.92\times10^{-6}$ \\
$N$ & Within-dist. & Stress $\delta=0.25$ & $0.7022$ & $0.9805$ & $-0.2783$ & $2.57\times10^{-126}$ & $\beta=-2.82\times10^{-4}$, $p=1.82\times10^{-21}$ \\
$N$ & OOD-high     & No stress            & $0.9136$ & $0.7961$ & $+0.1175$ & $1.74\times10^{-18}$ & $\beta=-1.50\times10^{-4}$, $p=0.0362$ \\
$N$ & OOD-high     & Stress $\delta=0.25$ & $0.6353$ & $0.8186$ & $-0.1833$ & $4.69\times10^{-50}$ & $\beta=-1.82\times10^{-4}$, $p=0.00125$ \\
\hline
\end{tabular}
\caption*{
Mean accuracies are pooled across all points within each sweep and reported for naive and resilient networks. Each sweep point contains 50 independently trained networks per class. The accuracy difference is defined as $\Delta^a=\bar a_{\mathrm{Naive}}-\bar a_{\mathrm{Resilient}}$, such that positive values favour naive networks. Group comparisons use independent two-sided Welch $t$-tests on the pooled network-level accuracies. The interaction column reports the network-type-by-parameter coefficient and its $p$-value from an ordinary least-squares model with HC3 standard errors, where the parameter is $\delta_T$ for the stress-training sweep and $N$ for the network-size sweep. The interaction coefficient represents the resilient slope minus the naive slope and tests whether the between-class accuracy difference changed linearly across the swept parameter.}
\end{table}

\end{document}